\documentclass[12pt]{article}

\usepackage[utf8]{inputenc}
\usepackage{latexsym, graphicx} 
\usepackage{amsmath,amsfonts,amssymb}
\usepackage{slashed}
\usepackage{cancel}
\usepackage{mathrsfs}
\usepackage{tensor}
\usepackage{xcolor}
\usepackage{braket}
\usepackage{cite}
\usepackage[colorlinks=true,linkcolor=blue,citecolor=blue,urlcolor=blue,filecolor=black]{hyperref}

\renewenvironment{thebibliography}[1]{
	\begin{oldthebibliography}{#1}
		\setlength{\itemsep}{2pt}
		\setlength{\parskip}{2pt}
	}
	{
	\end{oldthebibliography}
}

\numberwithin{equation}{section} 

\newcommand*{\scri}{\ensuremath{\mathscr{I}}}

\newcommand*{\dd}{\mathop{}\!d}
\newcommand*{\sgn}{\text{sgn}}
\newcommand*{\zbar}{\bar{z}}

\newcommand*{\zb}{\bar{z}}
\newcommand*{\hb}{\bar{h}}
\newcommand*{\nb}{\bar{n}}
\newcommand*{\kb}{\bar{k}}
\newcommand*{\Lb}{\bar{L}}

\newcommand*{\x}{\textbf{x}}

\begin{document}
	
	\begin{titlepage}
		\thispagestyle{empty}
		
		\begin{flushright}
		\end{flushright}
		
		\vskip1cm
		
		\begin{center}  
			{\Large\textbf{Operator Product Expansion in Carrollian CFT}}
			
			\vskip1cm
			
			\centerline{Kevin Nguyen and Jakob Salzer}
			
			\vskip1cm
			
			{\it{Universit\'e Libre de Bruxelles and International Solvay Institutes,\\ ULB-Campus Plaine
					CP231, 1050 Brussels, Belgium}}\\
			\vskip 1cm
			{kevin.nguyen2@ulb.be, jakob.salzer@ulb.be}
			
		\end{center}
		
		\vskip1cm
		
		\begin{abstract} 
			Carrollian conformal field theory offers an alternative description of massless scattering amplitudes, that is holographic in nature. In an effort to build a framework that is both predictive and constraining, we construct operator product expansions (OPE) that are compatible with carrollian symmetries. In this way, we unify and extend preliminary works on the subject, and demonstrate that the carrollian OPEs indeed control the short-distance expansion of carrollian correlators and amplitudes. In the process, we extend the representation theory of carrollian conformal fields such as to account for composite operators like the carrollian stress tensor or those creating multiparticle states. In addition we classify 2- and 3-point carrollian correlators and amplitudes with complex kinematics, and give the general form of the 4-point function allowed by symmetry.
		\end{abstract}
		
	\end{titlepage}
	
	{\hypersetup{linkcolor=black}
		\tableofcontents 
	}
	
	\section{Introduction}
	\label{section intro}
	The program of \textit{carrollian holography} aims at providing a holographic
	description of quantum gravity in asymptotically flat spacetimes in
	terms of a conformal field theory (CFT) defined on the spacetime null conformal
	boundary $\scri \cong \mathbb{R} \times \mathbb{CS}^2$, called \textit{carrollian} CFT \cite{Bagchi:2016bcd,Banerjee:2018gce,Donnay:2022aba,Bagchi:2022emh}. Within
	this approach the full BMS group
	\cite{Bondi:1962px,Sachs:1962zza,Sachs:1962wk} and its Poincaré
	subgroup act as conformal isometries of $\scri$, which
	has allowed in particular to interpret massless scattering amplitudes
	as a set of correlators of a carrollian CFT
	\cite{Banerjee:2019prz,Banerjee:2020kaa,Donnay:2022wvx,Bagchi:2023fbj,Saha:2023hsl,Salzer:2023jqv,Nguyen:2023vfz,Saha:2023abr,Nguyen:2023miw,Bagchi:2023cen,Mason:2023mti,Chen:2023naw,Alday:2024yyj,Stieberger:2024shv,Kraus:2024gso,Banerjee:2024hvb,Kraus:2025wgi}. However there is currently no intrinsic definition of what a carrollian CFT really is beyond simple kinematics, nor any toolbox which would allow to compute and predict correlators within a given carrollian CFT. This current state of affairs severely limits the usefulness of this program, as it does not yield new results about quantum gravity or even scattering theory. In this work we aim to bridge this gap by defining a new inherent structure of carrollian CFTs, the \textit{operator product expansion} (OPE). Just as in standard conformal field theory, its existence would place nontrivial constraints on the spectrum and interactions of a given theory, while at the same time allowing one to compute higher-point functions from the knowledge of lower-point functions. 
	
	The discussion around the existence of a conformal carrollian OPE is not entirely new. Indeed the authors of \cite{Banerjee:2020kaa} initiated its study, based on particular examples of carrollian correlators corresponding to specific massless scattering amplitudes. In this paper, we will generalize, correct, and complete their initial results. Our approach will be entirely self-contained, building consistent OPE structures requiring consistency with the action of $\operatorname{ISO}(1,3)$ viewed as conformal isometries of $\scri$, and subsequently testing their relevance and validity on explicit examples of carrollian correlators and amplitudes. Moreover, it has been argued that the colinear factorization of tree-level massless scattering amplitudes implies the existence of a different carrollian OPE for the corresponding carrollian correlators \cite{Mason:2023mti}. Our work will unify these various OPEs within a single framework. 
	
	Even though we are able to provide a fully intrinsic discussion of a carrollian OPE without having to invoke carrollian holography, it should be mentioned that this discussion is, for the moment, rather formal. To the best of the authors' knowledge, there exists at this moment no fully explicit example of an interacting, three-dimensional, quantum conformal carrollian theory. One could blame this on the requirement of having a concrete example of a conformal theory, which are also sparse in the standard CFT set-up. However, it turns out that this difficulty also extends to non-conformal quantum carrollian theories, which exhibit a number of surprising features \cite{deBoer:2023fnj,Cotler:2024xhb}. Such (non-conformal) carrollian theories \cite{Duval:2014uoa,Henneaux:2021yzg,deBoer:2021jej,Sharma:2025rug} have recently found increasing interest because they are expected to describe physics on generic null surfaces \cite{Teitelboim:1972vw,Henneaux:1979vn}, and serve as concrete examples of exotic (non-Lorentzian) quantum field theories closely related to fractons \cite{Bidussi:2021nmp, Baig:2023yaz,Kasikci:2023tvs}. This sparseness of concrete working examples of (conformal) carrollian quantum theories appears therefore to be one of the major challenges for the program of carrollian holography described in the preceding paragraphs. In the absence of toy models the approach adopted here is that of the `bootstrap', i.e., proceeding by imposing necessary consistency conditions in order to isolate candidate observables of a consistent theory, if it exists.

	The paper is organised as follows. In Section~\ref{section 2} we review some basic aspects of carrollian conformal field theory, starting with the realisation of $\operatorname{ISO}(1,3)$ as a subgroup of conformal isometries of $\scri \simeq \mathbb{R} \times S^2$. We also review the construction of carrollian field representations carrying massless one-particle states, inducing the full representations from representations of the isotropy subgroup of $\scri$ \cite{Nguyen:2023vfz}. We then generalize the method to build reducible but indecomposable carrollian fields, among which the carrollian stress tensor multiplet of \cite{Donnay:2022aba} as well as new \textit{massive} carrollian fields which we think might describe multi-particle states. In Section~\ref{section 3} we build a catalogue of 2-, 3-, and 4-point carrollian correlators of complex kinematics. These are essential to the carrollian description of massless scattering amplitudes, since the only nontrivial three-point amplitudes in the sense of tempered distributions are (anti)holomorphic functions. As is well-known, the form of these low-point correlators is not entirely determined by symmetries ; there are various `branches' of carrollian correlators. In Section~\ref{section 4} we focus on the set of carrollian correlators corresponding to massless tree-level MHV amplitudes, which will be used in later sections to test the validity of our carrollian OPEs. In Section~\ref{section 5} we come to the core of this work, namely the construction of the carrollian OPEs that are consistent with $\operatorname{ISO}(1,3)$ symmetries. Just as there are various branches of low-point correlators, there are also various branches of carrollian OPEs, which we will investigate with as much generality as we reasonably can. We start in Section~\ref{section 5.1} by studying the carrollian OPE in the uniform coincidence limit $\x_{12} \to 0$ where the operators $O_1(\x_1)O_2(\x_2)$ collide. We then consider the holomorphic coincidence limit $z_{12} \to 0$ in Section~\ref{section 5.2}, in order to discuss the OPE derived in \cite{Mason:2023mti} in relation to colinear factorisation of massless scattering amplitudes. We end in Section~\ref{section 5.3} by constructing OPE blocks valid at finite separations $\x_{12}\neq 0$, motivated by similar constructions in standard conformal field theory \cite{Czech:2016xec,Guevara:2021tvr}, and show how these relate to the carrollian OPEs built in Section~\ref{section 5.1}. In Section~\ref{section 6} we investigate whether the carrollian OPEs constructed ab initio are actually realised in practice, by looking at the coincide limit of the carrollian correlators and amplitudes listed in Sections~\ref{section 3}-\ref{section 4}. In all cases, the coincidence limit is consistent with the exchange of carrollian primary operators whose quantum numbers are determined, and the corresponding OPEs. In this way we gather substantial evidence that the carrollian OPEs constructed here are part of the defining structure of carrollian CFTs. As an illustration of the usefulness of the carrollian OPE, we show in particular that 3-point carrollian amplitudes are fully determined from a single OPE block and the knowlegde of 2-point amplitudes. 
	
	\section{Carrollian conformal field theory}
	\label{section 2}
	The carrollian conformal field theory that we will consider in this work mostly concerns massless particles as defined by Wigner, i.e., unitary irreducible representations of the Poincar\'e group $\operatorname{ISO}(1,3)$ with vanishing quadratic Casimir invariant \cite{Wigner:1939cj}. Because the Poincar\'e group is realised as a group of conformal isometries of the carrollian manifold $\scri \approx \mathbb{R} \times S^2$, viewed as the homogeneous space
	\cite{Herfray:2020rvq,Figueroa-OFarrill:2021sxz}
	\begin{equation}
	\scri\simeq \frac{\operatorname{ISO}(1,3)}{(\operatorname{ISO}(2) \ltimes \mathbb{R}^3) \rtimes \mathbb{R}}\,,
	\end{equation}
	one can construct conformal field representations of $\operatorname{ISO}(1,3)$ living on $\scri$ which encode the massless particle states \cite{Banerjee:2018gce,Nguyen:2023vfz}. This manifold is endowed with a conformal equivalence class of degenerate metrics, with standard representative
	\begin{equation}
	\label{scri metric}
	ds^2_{\scri}=0\, \dd u^2+\delta_{ij} \dd x^i \dd x^j=0\, \dd u^2+\dd z \dd \zbar\,.
	\end{equation}
	Here $x^i=(x^1,x^2)$ are cartesian stereographic coordinates on the sphere $S^2$ and $(z,\zbar)=(x^1+i x^2,x^1-i x^2)$ are complex ones. We will denote the full set of coordinates by $\x=(u,x^i)$. Of particular interest is the realisation of $\scri$ as the future or past component of the null conformal boundary of Minkowski spacetime. See \cite{Nguyen:2022zgs,Nguyen:2020hot} and references therein for a geometrical description of $\scri$ in the context of asymptotically flat gravity. The conformal transformations generating the Poincar\'e group can be explicitly written,
	\begin{align}
	\label{finite conformal transformaions}
	\begin{split}
	x'^i&=x^i+a^i\,, \hspace{6cm} (\text{spatial translations})\,,\\
	x'^i&=\Lambda\indices{^i_j} x^j\,, \hspace{7.2cm} (\text{rotations})\,,\\
	x'^i&=\lambda x^i\,, \quad u'=\lambda u\,, \hspace{5.6cm} (\text{dilation})\,,\\
	u'&=u+a^u\,, \hspace{6.3cm} (\text{time translation})\,,\\
	u'&=u+b_i x^i\,, \hspace{6.3cm} (\text{carroll boosts})\,,\\
	x'^i&=\frac{x^i-k^i\, x^2}{1-2\,k\cdot x+k^2\, x^2}\,, \quad u'=\frac{u-k^u\, x^2}{1-2\,k\cdot x+k^2\, x^2}\,, \quad (\text{SCT})\,.
	\end{split}
	\end{align}
	This can be easily obtained by taking the ultra-relativistic limit of the standard $\operatorname{SO}(2,3)$ conformal transformations of $\mathbb{M}^3$, following the method described in appendix A of \cite{Nguyen:2023vfz}.
	We note that the transformations of the spatial coordinates $x^i$ are the usual conformal transformations in $\mathbb{R}^2$, which means in particular that the usual conformal cross ratios built out of the spatial coordinates are invariant. Going to complex coordinates we can equivalently write \eqref{finite conformal transformaions} as
	\begin{align}
	\label{complex coordinate transformations}
	\begin{split}
	z'&=z+a\,, \hspace{7cm} (\text{spatial translations})\,,\\
	z'&=e^{i\theta} z\,, \hspace{8.1cm} (\text{rotation})\,,\\
	z'&=\lambda z\,, \quad u'=\lambda u\,, \hspace{6.3cm} (\text{dilation})\,,\\
	u'&=u+a^u\,, \hspace{7cm} (\text{time translation})\,,\\
	u'&=u+b \zbar+\bar b z\,, \hspace{6.3cm} (\text{carroll boosts})\,,\\
	z'&=\frac{z-k z\zbar}{1-k \zbar-\bar k z+k\bar k z\zbar}\,, \quad u'=\frac{u-k^u z\zbar}{1-k \zbar-\bar k z+k\bar k z\zbar}\,, \quad (\text{SCT})\,,
	\end{split}
	\end{align}
	together with the complex conjugate relations. 
	Given the standard basis of Poincar\'e generators $\langle \tilde J_{\mu\nu}, \tilde P_\mu \rangle$, the algebra elements generating the infinitesimal transformations parametrised by $(a^i,a^u, \Lambda^{ij},\lambda,b^i,k^i,k^u)$ respectively are $\langle P_i,H,J_{ij},D,B_i,K_i,K \rangle$, defined via
	\begin{equation}
	\label{tilde J}
	\tilde J_{ij}=J_{ij}\,, \qquad \tilde J_{i0}=-\frac{1}{2}\left(P_i+K_i \right)\,, \qquad \tilde J_{i3}= \frac{1}{2}\left(P_i-K_i \right)\,, \qquad \tilde J_{03}=-D\,,
	\end{equation}
	and
	\begin{equation}
	\label{tilde P}
	\tilde P_0=\frac{1}{\sqrt{2}}(H+K)\,, \qquad \tilde P_i=-\sqrt{2}\, B_i\,, \qquad \tilde P_3=\frac{1}{\sqrt{2}}(K-H)\,.
	\end{equation} 
	With respect to this alternative basis, the Poincar\'e algebra explicitly reads
	\begin{align}
	\nonumber
	\left[J_{ij}\,,J_{mn}\right]&=-i\left(\delta_{im} J_{jn}+\delta_{jn} J_{im}-\delta_{in} J_{jm}-\delta_{jm} J_{in} \right)\,, & \left[D\,, P_i \right]&=i P_i\,,\\
	\nonumber
	\left[J_{ij}\,,P_k \right]&=-i \left(\delta_{ik} P_j-\delta_{jk} P_i \right)\,, & \left[D\,, H \right]&=i H\,,\\
	\nonumber
	\left[J_{ij}\,,K_k \right]&=-i \left(\delta_{ik} K_j-\delta_{jk} K_i \right)\,, & \left[D\,, K_i \right]&=-i K_i\,,\\
	\label{conformal Carroll algebra}
	\left[J_{ij}\,,B_k \right]&=-i\left(\delta_{ik}B_j-\delta_{jk} B_i\right)\,, & \left[D\,, K \right]&=-i K\,,\\
	\nonumber
	\left[B_i\,, P_j \right]&=i \delta_{ij} H\,, & \left[H\,, K_i\right]&=2i B_i\,,\\
	\nonumber
	\left[B_i\,,K_j \right]&=i \delta_{ij} K\,, & \left[K\,, P_i\right]&=2i B_i\,,\\
	\nonumber
	\left[K_i\,, P_j\right]&=-2i\left(\delta_{ij} D-J_{ij} \right)\,,
	\end{align}
	with the remaining commutators being zero.
	
	Having introduced the basic kinematic ingredients, we can turn to the description of field representations of the Poincar\'e algebra, more specifically field representations living at~$\scri$. These can be constructed using the method of induced representations, starting from a representation of the isotropy subgroup of $\scri$, and subsequently `translating' using the remaining group elements. This method was used in \cite{Nguyen:2023vfz} to construct carrollian conformal fields encoding massless one-particle states. We first recall this construction, and then turn to the construction of more exotic field representations that should play a role in describing `composite' operators, such as two-particle states or the carrollian energy-momentum tensor. These new representations have the unusual feature of being reducible but indecomposable.  
	
	\subsection{One-particle fields}
	The carrollian conformal fields $O_{\Delta,J}(\x)$ carrying massless one-particle states are labeled by a scaling dimension $\Delta$ and a spin-$s$ representation of the massless little group. Their construction \cite{Nguyen:2023vfz} follows the method of induced representations, starting from a representation of the isotropy group of $\scri$ characterised by
	\begin{equation}
	\label{O little group}
	[J_{ij}\,,O_{\Delta,J}]=\Sigma^{(s)}_{ij}\, O_{\Delta,J}\,, \quad [D\,,O_{\Delta,J}]=i \Delta_{\phi}\, O_{\Delta,J}\,, \quad [B_i\,,O_{\Delta,J}]=[K_\alpha\,,O_{\Delta,J}]=0\,.
	\end{equation}
	We then induce the full Poincar\'e representation by `translating' the fields,
	\begin{equation}
	\label{translated fields}
	O_{\Delta,J}(\x)\equiv U(\x)\, O_{\Delta,J}\, U(\x)^{-1}\,, 
	\end{equation}
	with the group elements
	\begin{equation}
	U(\x)=e^{-i x^\alpha P_\alpha}=e^{-i(uH+x^i P_i)}\,.
	\end{equation}
	This definition directly implies
	\begin{equation}
	[P_\alpha\,,O_{\Delta,J}(\x)]=i\partial_\alpha O_{\Delta,J}(\x)\,.
	\end{equation}
	To work out the action of one of the isotropy generators $X$ on the translated field $O_{\Delta,J}(\x)$, we make use of the identity
	\begin{equation}
	[X\,,\psi_i(\x)]=U(\x) [X'\,,\psi_i\,] U(\x)^{-1}\,,    
	\end{equation}
	where
	\begin{equation}
	X'=U(\x)^{-1}XU(\x)=\sum_{n=0}^\infty \frac{i^n}{n!} x^{\alpha_1}...\,x^{\alpha_n}\, [P_{\alpha_1}\,,[...\,,[P_{\alpha_n}\,,X]]]\,.
	\end{equation}
	Using the algebra relations, we explicitly evaluate the primed generators, given by
	\begin{equation}
	\label{translated generators}
	\begin{split}
	J_{ij}'&=J_{ij}-x_i P_j+x_j P_i\,,\\
	D'&=D+uH+x^iP_i\,,\\
	K'&=K+2x^i B_i+x^2 H\,,\\
	K_i'&=K_i-2uB_i-2x_i D+2x^jJ_{ij}-2x_iuH-2x_ix^jP_j+x^2P_i\,,\\
	B_i'&=B_i+x_iH\,.
	\end{split}
	\end{equation}
	Hence we have, for instance,
	\begin{equation}
	\begin{split}
	[D\,,O_{\Delta,J}(\x)]&=U(\x) [D+x^\alpha P_\alpha\,,O_{\Delta,J}] U(\x)^{-1}\\
	&=U(\x) \left(i\Delta\, O_{\Delta,J}+x^\alpha[P_\alpha\,,O_{\Delta,J}]\right) U(\x)^{-1}\\
	&=i\left(\Delta+x^\alpha\partial_\alpha\right)O_{\Delta,J}(\x)\,,\\
	\end{split}
	\end{equation}
	where in the last line we made use of
	\begin{equation}
	\begin{split}
	U(\x)[P_\alpha\,,O_{\Delta,J}]U(\x)^{-1}&=U(\x)P_\alpha\, O_{\Delta,J} U(\x)^{-1}-U(\x) O_{\Delta,J} P_\alpha U(\x)^{-1}\\
	&=P_\alpha\, U(\x) O_{\Delta,J} U(\x)^{-1}-U(\x) O_{\Delta,J} U(\x)^{-1} P_\alpha\\
	&=[P_\alpha\,,O_{\Delta,J}(\x)]=i\partial_\alpha O_{\Delta,J}(\x)\,.
	\end{split}
	\end{equation}
	Similar manipulations can be performed for the remaining generators, yielding \cite{Nguyen:2023vfz}
	\begin{equation}
	\label{Carrollian induced rep}
	\begin{split}
	\left[P_\alpha, O_{\Delta,J}(\x)\right]&=i \partial_\alpha O_{\Delta,J}(\x)\,,\\
	\left[J_{ij}, O_{\Delta,J}(\x)\right]&=i(-x_i \partial_j+x_j\partial_i -i\Sigma^{(s)}_{ij})\,O_{\Delta,J}(\x)\,,\\
	\left[D, O_{\Delta,J}(\x)\right]&=i(\Delta+u\partial_u+x^i\partial_i )\,O_{\Delta,J}(\x)\,,\\
	\left[K, O_{\Delta,J}(\x)\right]&=i x^2 \partial_u O_{\Delta,J}(\x)\,,\\
	\left[K_i, O_{\Delta,J}(\x)\right]&=i(-2x_i \Delta-2ix^j \Sigma_{ij}-2x_i u\partial_u-2x_i x^j \partial_j+x^j x_j \partial_i )\,O_{\Delta,J}(\x)\,,\\
	\left[B_i, O_{\Delta,J}(\x)\right]&=ix_i\partial_u O_{\Delta,J}(\x)\,.
	\end{split}
	\end{equation}
	While these apply to arbitrary dimension, for $\operatorname{ISO}(1,3)$ a spin-$s$ representation of the little group $\operatorname{SO}(2)$ breaks into two helicity components $O_{\Delta,J}$ with helicity $J=\pm s$, such that we can write
	\begin{equation}
	\Sigma_{ij}^{(s)}\, O_{\Delta,J}=J \varepsilon_{ij}\, O_{\Delta,J}\,.
	\end{equation}
	These carrollian conformal fields can be related to the massless particle states $|p(\omega,x^i)\rangle_J$ of helicity $J$ and momentum $p^\mu$ parametrised as
	\begin{equation}
	\label{momentum parametrisation}
	p^\mu(\omega,x^i)=\frac{\omega}{\sqrt{2}}(1+x^2,2x^i,1-x^2)\,, 
	\end{equation}
	through the \textit{modified Mellin transform} \cite{Banerjee:2018gce,Banerjee:2019prz,Donnay:2022wvx,Nguyen:2023vfz}
	\begin{equation}
	\label{Mellin}
	O_{\Delta,J}(u,x^i)|0\rangle =\int_0^\infty d\omega\, \omega^{\Delta-1} e^{i\omega u} |p(\omega,x^i)\rangle_J\,.    
	\end{equation}
	This provides a simple intertwining relation between the particle states of a unitary theory and the carrollian conformal fields. Note however that even though carrollian conformal fields can be realised in this way, they are in fact more general objects. In particular their correlation functions can be more general than those obtained by applying the modified Mellin transform \eqref{Mellin} to generic $\mathcal{S}$-matrix elements \cite{Nguyen:2023miw}. 
	
	When working in complex coordinates $(z,\zbar)=(x_1+ix_2,x_1-i x_2)$, it is more natural to define the generators~\cite{Bagchi:2023cen}
	\begin{equation}
	\begin{aligned}
	P_{-1,-1}&=-iH\,, &\quad  L_{-1}&=-\frac{i}{2}(P_1-iP_2)\,, &\quad  \bar L_{-1}&=-\frac{i}{2}(P_1+iP_2)\,,\\
	P_{0,-1}&=-i(B_1+iB_2)\,, &\quad L_0&=-\frac{i}{2}(D+iJ_{12})\,, &\quad \bar L_0&=-\frac{i}{2}(D-iJ_{12})\,,\\
	P_{-1,0}&=-i(B_1-iB_2)\,, &\quad L_1&=\frac{i}{2}(K_1+iK_2)\,, &\quad \bar L_1&=\frac{i}{2}(K_1-iK_2)\,,\\
	P_{0,0}&=-iK
	\end{aligned}
	\end{equation}
	such that we can write 
	\begin{align}
	\label{complex field transformations}
	\begin{split}
	\left[P_{-1,-1}\,,O_{\Delta,J}(\x) \right]&=\partial_u O_{\Delta,J}(\x)\,,\\
	\left[P_{0,-1}\,,O_{\Delta,J}(\x)\right]&=z\partial_u O_{\Delta,J}(\x)\,,\\
	\left[P_{-1,0}\,,O_{\Delta,J}(\x)\right]&=\zbar \partial_u O_{\Delta,J}(\x)\,,\\
	\left[P_{0,0}\,,O_{\Delta,J}(\x)\right]&=z\zbar \partial_u O_{\Delta,J}(\x)\,,
	\end{split}
	\end{align}
	together with
	\begin{align}
	\begin{split}
	\left[L_{-1}\,,O_{\Delta,J}(\x)\right]&=\partial_z O_{\Delta,J}(\x)\,,\\
	\left[L_0\,,O_{\Delta,J}(\x)\right]&=\frac{1}{2} \left(u\partial_u+2z\partial_z+2h \right)O_{\Delta,J}(\x)\,,\\
	\left[L_1\,,O_{\Delta,J}(\x) \right]&= z\left(u\partial_u+z\partial_z+2h \right)O_{\Delta,J}(\x)\,,
	\end{split}
	\end{align}
	and the conjugate relations, where we defined the chiral weights 
	\begin{equation}
	\label{h hbar}
	h=\frac{\Delta+J}{2}\,, \qquad \bar h=\frac{\Delta-J}{2}\,.
	\end{equation}
	Note that one recovers the standard $SL(2,\mathbb{C})$ conformal field transformations by imposing $\partial_u O=0$, which ensures that the abelian translations $\tilde P_\mu=\{H,K,B_i\}$ act trivially. In that case $h,\bar h$ are the usual conformal weights.
	When $\Delta=1$ the transformations \eqref{complex field transformations} agree with those of \cite{Donnay:2022wvx} upon identifying $O_{z...z}=O_{J=s}$ and $O_{\zbar...\zbar}=O_{J=-s}$. Furthermore it can be seen that the transformation above are the infinitesimal version of
	\begin{equation}
	\label{conformal transfo}
	O'_{\Delta,J}(\x')=\left(\frac{\partial z'}{\partial z}\right)^{-h} \left(\frac{\partial \zbar'}{\partial \zbar}\right)^{-\bar h} O_{\Delta,J}(\x)\,.
	\end{equation}
	
	In fact the Poincar\'e group is only a subgroup of the conformal group of $\scri$, known as the (extended) BMS group \cite{Barnich:2010eb}. Indeed we can consider the set of generators $\{L_n\,, \bar L_n\,, P_{m,n}\}$ with commutation relations
	\begin{equation}
	\label{[L,P]}
	[L_n,P_{m,k}]=\left(\frac{n-1}{2}-m\right)P_{m+n,k}\,, \qquad [\bar{L}_n,P_{m,k}]=\left(\frac{n-1}{2}-k\right)P_{m,k+n}\,,
	\end{equation}
	and 
	\begin{equation}
	\label{[L,L]}
	[L_m,L_n]=(m-n)L_{m+n}\,, \qquad [\bar{L}_m,\bar{L}_n]=(m-n)\bar{L}_{m+n}\,.
	\end{equation}
	An extension of the transformations \eqref{complex field transformations} which realises this algebra is given by \cite{Bagchi:2016bcd,Banerjee:2020kaa}
	\begin{equation}
	\label{eq:supertrans}
	\begin{split}
	[P_{m,n}\,, O_{\Delta,J}(\x)]&=z^{m+1}\zbar^{n+1} \partial_u O_{\Delta,J}(\x)\,,\\
	[L_n\,,O_{\Delta,J}(\x)]&=(z^{n+1}\partial_z+(n+1)(h+\frac{1}{2}u\partial_u)z^n)\,O_{\Delta,J}(\x)\,,
	\end{split}
	\end{equation}
	together with the conjugate relations. Finally, we note that we can also `translate' the descendants of a primary operator. For a generic element $G$ in the enveloping algebra of the BMS group, we define
	\begin{equation}
	\label{descendant fields}
	(GO_{\Delta,J})(\x)\equiv U(\x) [G\,,O_{\Delta,J}] U(\x)^{-1}\,.
	\end{equation}
	According to this definition and the action \eqref{eq:supertrans} of the BMS generator on $O_{\Delta,J}\equiv O_{\Delta,J}(0)$, the nonzero descendant fields of degree one are the supertranslation descendants $(P_{m,n}O_{\Delta,J})(\x)$ for $m,n\leq -1$, in addition to the familiar Virasoro descendants $(L_n O_{\Delta,J})(\x)$ for $n\leq -1$. Just as in standard conformal field theory, the descendant fields will appear in the operator product expansion.  
	
	\subsection{Carrollian stress tensor multiplet}
	In this subsection and the next, we discuss some reducible but indecomposable carrollian field representations. These representations should be thought of `composite' operators arising from products of fundamental massless particles. We will follow the general procedure of induced representations, starting from a representation of the isotropy group of $\scri$ which will itself be indecomposable. More specifically we will consider a pair $(\phi\,, \psi)$, with $\phi$ the irreducible component transforming according to the isotropy group as 
	\begin{equation}
	\label{phi little group}
	[J_{ij}\,,\phi]=\Sigma_{ij}\, \phi\,, \quad [D\,,\phi]=i \Delta_{\phi}\, \phi\,, \quad [B_i\,,\phi]=[K_\alpha\,,\phi]=0\,,
	\end{equation}
	just like in the case of irreducible one-particle fields.
	The remaining component $\psi$ will not transform autonomously however.
	
	We aim to describe the carrollian multiplet which will account for the BMS charge aspects constructed in \cite{Donnay:2022hkf,Barnich:2022bni}, that can also be viewed to be the components of a carrollian stress tensor \cite{Chandrasekaran:2021hxc,Donnay:2022aba,Bagchi:2024gnn,Ruzziconi:2024kzo,Ciambelli:2025mex}. Here we discuss it from the perspective of representation theory. For this we consider the irreducible component $\phi$ to be a carrollian conformal scalar ($[J_{ij},\phi]=0$) and $\psi\equiv \psi_i$ a representation of the isotropy group satisfying
	\begin{equation}
	[J_{ij}\,,\psi_k]=(\Sigma_{ij})\indices{_k^l}\, \psi_l\,, \quad [D\,, \psi_i]=i\Delta_{\psi}\, \psi_i\,, \quad [B_i\,,\psi_j]=i\delta_{ij}\, \phi\,, \quad [K_\alpha,\psi_i]=0\,,
	\end{equation}
	with $(\Sigma_{ij})\indices{_{kl}}=i(\delta_{il}\delta_{jk}-\delta_{ik}\delta_{jl})$ the vector representation of $SO(2)$. The unusual transformation is the one generated by carroll boosts $B_i$. Of course it is necessary to impose consistency with all commutation relations of the isotropy subgroup. In particular we compute
	\begin{equation}
	\begin{split}
	[[D\,,B_i]\,,\psi_j]&=[D\,,[B_i\,,\psi_j]]-[B_i\,,[D\,,\psi_j]]=i \delta_{ij}\, [D\,,\phi]-i\Delta_\psi\, [B_i\,,\psi_j]\\
	&=\delta_{ij}(\Delta_\psi-\Delta_\phi)\phi\,,\\
	\end{split}
	\end{equation}
	such that we have to impose 
	\begin{equation}
	\Delta_\psi=\Delta_\phi\,.
	\end{equation}
	It is also interesting to compute the action of the quadratic Casimir operator on the `exotic' component $\psi_i$, giving
	\begin{equation}
	[\mathcal{C}_2\,,\psi_i]=-2[H\,,[K\,,\psi_i]]+2[B^j\,,[B_j\,,\psi_i]]=2i[B_i\,,\phi]=0\,.
	\end{equation}
	This shows that the representation $(\phi,\psi_i)$ is a massless representation.
	We then induce the full Poincar\'e representation by `translating' the fields as in \eqref{translated fields},
	\begin{equation}
	\label{translated fields}
	\phi(\x)\equiv U(\x)\, \phi\, U(\x)^{-1}\,, \qquad \psi_i(\x)\equiv U(\x)\, \psi_i\, U(\x)^{-1}\,,
	\end{equation}
	and working out the resulting symmetry transformations. Those of $\phi(\x)$ are given by \eqref{Carrollian induced rep} with $\Sigma_{ij}=0$, while for $\psi_i(\x)$ we obtain
	\begin{equation}
	\label{transformation psi_i}
	\begin{split}
	[P_\alpha\,,\psi_i(\x)]&=i\partial_\alpha \psi_i(\x)\,,\\
	[J_{ij}\,,\psi_k(\x)]&=i(-x_i \partial_j+x_j \partial_i-i\Sigma_{ij})\psi_k(\x)\,,\\
	[D\,,\psi_i(\x)]&=i(\Delta_\psi+x^\alpha\partial_\alpha)\psi_i(\x)\,,\\
	[K\,,\psi_i(\x)]&=i x^2\partial_u \psi_i(\x)+2ix_i\, \phi(\x)\,,\\ 
	[K_i\,,\psi_j(\x)]&=i(-2x_i \Delta_{\psi}-2x_i x^\alpha \partial_\alpha+x^2\partial_i-2i x^k\Sigma_{ik})\psi_j(\x) -2iu\delta_{ij}\,\phi(\x)\,,\\
	[B_i\,,\psi_j(\x)]&=ix_i\partial_u \psi_j(\x)+i\delta_{ij} \phi(\x)\,.
	\end{split}
	\end{equation}
	In complex coordinates, and going to helicity basis
	\begin{equation}
	\psi_{J=1}\equiv \psi_z=\frac{\psi_1-i\psi_2}{2}\,, \qquad \psi_{J=-1}\equiv \psi_{\zbar}=\frac{\psi_1+i\psi_2}{2}\,,
	\end{equation}
	they read
	\begin{align}
	\begin{split}
	\left[P_{-1,-1}\,,\psi_J(\x) \right]&=\partial_u \psi_J(\x)\,,\\
	\left[P_{0,-1}\,,\psi_J(\x)\right]&=z\partial_u \psi_J(\x)+\delta_{zJ}\,\phi(\x)\,,\\
	\left[P_{-1,0}\,,\psi_J(\x)\right]&=\zbar \partial_u \psi_J(\x)+\delta_{\zbar J}\, \phi(\x)\,,\\
	\left[P_{0,0}\,,\psi_J(\x)\right]&=z\zbar \partial_u \psi_J(\x)+\partial_J(z\zbar) \phi(\x) \,,
	\end{split}
	\end{align}
	together with
	\begin{align}
	\begin{split}
	\left[L_{-1}\,,\psi_J(\x)\right]&=\partial_z \psi_J(\x)\,,\\
	\left[L_0\,,\psi_J(\x)\right]&=\frac{1}{2} \left(u\partial_u+2z\partial_z+2h \right)\psi_J(\x)\,,\\
	\left[L_1\,,\psi_J(\x) \right]&= z\left(u\partial_u+z\partial_z+2h \right)\psi_J(\x)+u\delta_{zJ}\, \phi(\x)\,,
	\end{split}
	\end{align}
	and the conjugate relations.
	
It is interesting to compare this to the BMS transformations of the gravitational mass aspect $\mathcal{M}(\x)$ and Lorentz charge aspects $\mathcal{N}_i(\x)$ constructed in \cite{Donnay:2022hkf,Barnich:2022bni}. These are simply related to the Newman-Penrose Weyl scalars through \cite{Donnay:2022hkf}
\begin{equation}
\mathcal{M}=-\frac{1}{2}(\Psi^0_2+\bar{\Psi}^0_2)\,, \qquad \mathcal{N}_{\zbar}=-\Psi^0_1+u \partial_{\zbar} \Psi^0_2\,.    
\end{equation}
In absence of radiation, they transform as \cite{Barnich:2022bni}
\begin{equation}
\begin{split}
\delta \Psi^0_2(\x)&=( f \partial_u+Y\partial_z+\bar Y\partial_{\zbar}+\frac{3}{2} \partial_z Y+\frac{3}{2} \partial_{\zbar}\bar Y )\, \Psi^0_2(\x)\,,\\
\delta \Psi^0_1(\x)&=\left( f \partial_u+Y\partial_z+\bar Y\partial_{\zbar}+ \partial_z Y+ 2\partial_{\zbar}\bar Y \right)\Psi^0_1(\x)+ 3\partial_{\zbar} f\, \Psi^0_2(\x)\,,
\end{split}
\end{equation}
with 
\begin{equation}
f(\x)=T(z,\zbar)+\frac{u}{2}\left(\partial_z Y(z)+\partial_{\zbar} \bar Y(\zbar)\right)\,. 
\end{equation}
Extended BMS transformations consist of supertranslations parametrised by the function $T(z,\zbar)$ and superrotations parametrised by (anti)-holomorphic vector fields $Y(z)\, (\bar Y(\zbar))$. The Poincar\'e subgroup is generated by $T(z,\zbar)=\{1,z,\zbar,z \zbar\}$ that correspond to the translation generators $\{P_{-1,-1}\,, P_{0,-1}\,, P_{-1,0}\,, P_{0,0}\}$ and by $Y(z)=\{1\,,z\,,z^2\}$ corresponding to the Lorentz generators $\{L_{-1}\,,L_0\,,L_1\}$. It is easy to check that the BMS charge aspects transform exactly as $(\phi,\psi_i)$ under the Poincar\'e group, provided we set $\Delta_\phi=\Delta_\psi=3$ and we make the identifications
\begin{equation}
3\Psi^0_2=\phi\,, \qquad \Psi^0_1=\psi_{\zbar}\,, \qquad \bar{\Psi}^0_1=\psi_z\,.
\end{equation}
Note that here we assume $\Psi^0_2=\bar{\Psi}^0_2$ for simplicity, although one could also consider a complex carrollian field $\phi$.
Thus we have provided, from the perspective of carrollian conformal field theory developed here, the indecomposable representation corresponding to the BMS charge aspects. The authors of \cite{Donnay:2022aba} have argued that these make up the independent components of a carrollian stress tensor $T\indices{^\alpha_\beta}$ via
\begin{equation}
T\indices{^u_u}=2\Psi^0_2\,, \qquad T\indices{^u_\zbar}=\Psi^0_1\,.
\end{equation}
The defining property of this carrollian stress tensor is that it satisfies conservation equations in absence of radiation \cite{Donnay:2022aba}. 
	
	\subsection{Two-particle multiplets}
	\label{subsection two particle}
	Even though we are considering a theory composed of one-particle states which are strictly massless, it is interesting to wonder whether there exist massive representations that encode multiparticle states. In section~\ref{section 5} we will argue that these need to appear in a consistent OPE expansion. Inspired by the recent discussion in \cite{Kulp:2024scx}, in this subsection we construct a family of indecomposable carrollian conformal field representations, which we think could very well describe two-particle states. 
	We consider both $\phi$ and $\psi$ to have $\operatorname{SO}(2)$ spin, and we postulate the following isotropy transformations of the latter,
	\begin{equation}
	[D\,,\psi]=i\Delta_\psi\, \psi\,, \qquad [K,\psi]=H \phi\,, \qquad [K_i\,,\psi]=\kappa P_i\phi\,, \qquad [B_i\,,\psi]=i\beta P_i H \phi\,,    
	\end{equation}
	where we use the shorthand notation $G\phi\equiv [G,\phi]$.
	The parameter $\beta$ can be set to zero but we keep it for generality. Let us explicitly impose consistency with the commutation relations of the isotropy algebra. First aiming at checking $[D\,,K]=-i K$, we compute 
	\begin{equation}
	\begin{split}
	[[D\,,K]\,,\psi]&=[D\,,[K\,,\psi]]-[K\,,[D\,,\psi]]=[D\,,H\phi]-i\Delta_\psi [K\,,\psi]\\
	&=i(\Delta_\phi+1)H\phi-i\Delta_\psi\, H\phi=i(\Delta_\phi-\Delta_\psi+1) H\phi\,,\\
	\end{split}    
	\end{equation}
	and we thus require $\Delta_\psi=\Delta_\phi+2$. The commmutation $[D\,,K_i]=-i K_i$ is then automatically satisfied as well. We can also straightforwardly check
	\begin{equation}
	\begin{split}
	[[D\,,B_i]\,,\psi]&=[D\,,[B_i\,,\psi]]-[B_i\,,[D\,,\psi]]=i\beta\,[D\,,P_i H\phi]-i\Delta_\psi\, [B_i\,,\psi]\\
	&=\beta (\Delta_\psi-\Delta_\phi-2)P_iH\phi=0\,,\\
	[[K\,,K_i]\,,\psi]&=[K\,,[K_i\,,\psi]]-[K_i\,,[K\,,\psi]]=\kappa\, [K\,,[P_i\,,\phi]]-[K_i\,,H\phi]\\
	&=\kappa\, [[K\,,P_i]\,,\phi]=2\kappa\, [B_i\,,\phi]=0\,,\\
	[[K\,,B_i]\,,\psi]&=[K\,,[B_i\,,\psi]]-[B_i\,,[K\,,\psi]]=i\beta\, [K\,,[P_i\,,H\phi]]-[B_i\,,H\phi]\\
	&=i\beta\, [[K\,,P_i]\,,H\phi]=-2\beta\, [B_i\,,H\phi]=0\,.
	\end{split}
	\end{equation}
	To establish consistency with $[J_{ij}\,,K_k]=-i(\delta_{ik}K_j-\delta_{jk}K_i)$, we compute
	\begin{equation}
	\begin{split}
	[[J_{ij}\,,K_k]\,,\psi]&=[J_{ij}\,,[K_k\,,\psi]]-[K_k\,,[J_{ij}\,,\psi]]=\kappa\, [J_{ij}\,,[P_k\,,\phi]]-\Sigma_{ij}^\psi [K_k\,,\psi]\\
	&=\kappa\, [[J_{ij}\,,P_k]\,,\phi]+\kappa\, [P_k\,,[J_{ij}\,,\phi]]-\kappa\, \Sigma_{ij}^\psi P_k\phi\\
	&=-i\kappa\, (\delta_{ik}\, P_j\phi-\delta_{jk}\, P_i\phi)+\kappa\,(\Sigma_{ij}^\phi-\Sigma_{ij}^\psi)P_k\phi \,,
	\end{split}
	\end{equation}
	therefore requiring $\phi,\psi$ to have identical spin, $\Sigma_\psi=\Sigma_\phi$. 
	To check $[B_i\,,K_j]=i\delta_{ij} K$, we compute
	\begin{equation}
	\begin{split}
	[[B_i\,,K_j]\,,\psi]&=[B_i\,,[K_j\,,\psi]]-[K_j\,,[B_i\,,\psi]]=\kappa\, [B_i\,,[P_j\,,\phi]]-i\beta\, [K_j\,,[P_i\,,H\phi]]\\
	&=\kappa\, [[B_i\,,P_j]\,,\phi]-i\beta\, [ [K_j\,,P_i]\,,H\phi]=i\kappa\, \delta_{ij}\, H\phi-2\beta\, [\delta_{ij}D+J_{ij}\,,H\phi]\\
	&=i \delta_{ij}\left(\kappa-2\beta(\Delta_\phi+1)\right) H\phi-2\beta\, \Sigma_{ij}^\phi H\phi\,.
	\end{split}
	\end{equation}
	To cancel the second term we have to set $\beta=0$ unless $\phi$ is a scalar. In addition, we need to impose
	\begin{equation}
	\kappa=1+2\beta(\Delta_\phi+1)\,.
	\end{equation}
	In summary our representation $(\phi\,,\psi)$ of the isotropy subgroup is labeled by two free parameters $(\Delta_\phi\,,s_\phi)$ in the spinning case or $(\Delta_\phi\,,\beta)$ in the spinless case. 
	
	Having discussed the representation of the isotropy subgroup, we can induce the full representation using \eqref{translated generators}-\eqref{translated fields}. While $\phi(\x)$ transforms like a single-particle field, for $\psi(\x)$ we find 
	\begin{equation}
	\label{transformation psi 2-part}
	\begin{split}
	[P_\alpha\,,\psi(\x)]&=i\partial_\alpha \psi(\x)\,,\\
	[J_{ij}\,,\psi(\x)]&=i(-x_i \partial_j+x_j \partial_i-i\Sigma_{ij})\psi(\x)\,,\\
	[D\,,\psi(\x)]&=i(\Delta_\psi+x^\alpha\partial_\alpha)\psi(\x)\,,\\
	[K\,,\psi(\x)]&=i x^2\partial_u \psi(\x)+i( \partial_u-2\beta\, x^i\partial_i\partial_u ) \phi(\x)\,,\\ 
	[K_i\,,\psi(\x)]&=i(-2x_i \Delta_{\psi}-2x_i x^\alpha \partial_\alpha+x^2\partial_i-2i x^k\Sigma_{ik})\psi(\x)+i(\kappa\, \partial_i+2\beta\, u\partial_i\partial_u)  \phi(\x) \,,\\
	[B_i\,,\psi(\x)]&=ix_i\partial_u \psi(\x)-i\beta\, \partial_i \partial_u \phi(\x)\,,
	\end{split}
	\end{equation}
	where we recall $\Delta_\psi=\Delta_\phi+2$, and $\beta=0$ unless $\Sigma=\Sigma_\psi=\Sigma_\phi\neq 0$.
	At this point it is interesting to evaluate the action of the quadratic Casimir operator. While $\phi$ is massless by construction, for $\psi$ we find
	\begin{equation}
	\label{C2 psi}
	\begin{split}
	[\mathcal{C}_2\,,\psi(\x)]&=-2[H\,,[K\,,\psi(\x)]]+2[B^i\,,[B_i\,,\psi(\x)]]=2 (1+2\beta)\,\partial_u^2 \phi(\x)\,.
	%-2HH\phi+2i\beta\,\delta^{ij} [B_i\,,[P_j\,,H\phi]]\\
	%&=-2HH\phi+2i\beta\, \delta^{ij}\, [[B_i\,,P_j]\,,H\phi]=-2(1+\beta(d-1))HH\phi\,,
	\end{split}
	\end{equation}
	%where $d-1=2$ is the dimensionality of the celestial sphere $S^2$. 
	The above quantity is non-zero and $\psi(\x)$ thus has non-zero mass, unless $\phi(\x)$ is a zero-momentum representation satisfying $\partial_u \phi(\x)=0$. As we will discuss in section~\ref{section 5}, $\psi(\x)$ is the kind of operator we expect to see in the OPE of two single-particle operators.
	
	\section{Correlators of complex kinematics}
	\label{section 3}
	In any consistent theory whose vacuum is invariant under Poincar\'e symmetry, the correlators of the carrollian conformal fields must satisfy the Ward identity 
	\begin{equation}
	\label{Ward identities}
	\sum_{k=1}^n \langle O_1(\x_1)\,...\,\delta O_k(\x_k)\,...\,O_n(\x_n)\rangle =0\,,   
	\end{equation}
	with $\delta O$ any linear combinations of the Poincar\'e transformations, such as \eqref{Carrollian induced rep} if $O$ is a single-particle operator. We will indeed restrict our attention to correlators of single-particle operators because they account for scattering amplitudes. For real kinematics, i.e.~for $\zbar=z^*$, the corresponding two- and three-point functions solving the Ward identities have been classified in \cite{Nguyen:2023miw}. However to discuss massless scattering amplitudes it is essential to allow for complex kinematics where $z, \zbar$ are considered independent variables, since in particular the only nontrivial three-point amplitudes (in the sense of tempered distributions) are (anti)holomorphic functions. In this section we list the 2- and 3-point functions with complex kinematics that are solutions to the Ward identities \eqref{Ward identities}, and discuss the general form of the 4-point functions.   
	
	\subsection{2-point functions}
	It is known that two-point functions with real kinematics take the form \cite{Banerjee:2018gce,Bagchi:2022emh,Donnay:2022wvx,Nguyen:2023miw}
	\begin{equation}
	\label{2-point function 1}
	\langle O_1(\x) O_2(0)\rangle =a_{12}\, \frac{ \delta_{\Delta_1,\Delta_2}\,\delta_{J_1,J_2}}{|z|^{\Delta_1+\Delta_2}}+b_{12}\, \frac{ \delta(z)\delta(\zbar)\delta_{J_1,-J_2}}{u^{\Delta_1+\Delta_2-2}}\,,
	\end{equation}
	where the coefficients $a_{12},b_{12}$ can be arbitrary. We can find additional solutions to the Ward identities \eqref{Ward identities} if we allow for complex kinematics ($\zbar \neq z^*$), of the form
	\begin{equation}
	\langle O_1(\x) O_2(0) \rangle=f(u,z) \delta(\zbar)\,.    
	\end{equation}
	The Ward identities associated with $L_0$ and $\bar L_0$ yield
	\begin{align}
	\begin{split}
	\left[ u \partial_u+2z \partial_z+2(h_1+h_2) \right] f(u,z)&=0\,,\\  
	\left[ u \partial_u+2(\bar h_1+\bar h_2-1) \right] f(u,z)&=0\,,
	\end{split}
	\end{align}
	with solution
	\begin{equation}
	f(u,z)=u^{-2(\bar h_1+\bar h_2-1)}z^{-(J_1+J_2+1)}\,.
	\end{equation}
	While the Ward identities associated with $P_{-1,0}$ and $P_{0,0}$ are automatically satisfied, imposing the one of $P_{0,-1}$ requires $u$-independence, namely
	\begin{equation}
	\bar h_1+\bar h_2=1\,.
	\end{equation}
	Finally solving the Ward identity associated with $L_1, \bar L_1$ yields
	\begin{equation}
	h_1=h_2\,.
	\end{equation}
	Therefore the additional two-point function takes the form
	\begin{equation}
	\label{chiral 2-point function}
	\langle O_1(\x) O_2(0) \rangle =c_{12}\, \frac{\delta_{h_1,h_2}\,\delta_{\bar h_1+\bar h_2,1}\delta(\zbar)}{z^{2h_1}}  \,, 
	\end{equation}
	together with the conjugate solution. It is the product of a chiral two-point function in standard $\operatorname{CFT}_2$ with a singular anti-chiral two-point function.
	
	\subsection{3-point functions}
	\label{subsection 3.2}
	It is alos well-known that momentum conservation for three massless momenta requires all momenta to be colinear, i.e.,
	\begin{equation}
	\label{colinearity}
	p_1 \cdot p_2=p_2 \cdot p_3=p_1 \cdot p_3=0\,.
	\end{equation}
	In the momentum parametrisation \eqref{momentum parametrisation}, this reads
	\begin{equation}
	|z_{12}|^2=|z_{23}|^2=|z_{13}^2|=0\,.
	\end{equation}
	For real kinematics a nontrivial three-point distribution therefore contains a product of Dirac distributions $\delta^{(2)}(x_{12}^i)\delta^{(2)}(x_{23}^i)$ such as to make the above kinematic region give a nonzero contribution to momentum integrals \cite{Nguyen:2023miw}. 
	The corresponding three-point function takes the form \cite{Chang:2022seh,Bagchi:2023fbj,Nguyen:2023miw}
	\begin{equation}
	\label{eq:ultralocal3pt}
	\langle O_1O_2O_3\rangle=c_{123}\, \frac{\delta^{(2)}(x_{12}^i)\delta^{(2)}(x_{23}^i)}{(u_{12})^a (u_{23})^b (u_{31})^c},
	\end{equation}
	with 
	\begin{equation}
	a+b+c+4=\Delta_1+\Delta_2+\Delta_3.\qquad J_{1}+J_2+J_3=0.
	\end{equation}
	
	With complex kinematics we can have something less singular, of the form $\delta(\zbar_{12})\delta(\zbar_{23})$, which has become standard practice in the study of massless amplitudes. Here we aim to construct the general carrollian three-point function of this type. The first step is to find the quantities constructed out of the three coordinates $\x_1,\x_2,\x_3$ which are translation-invariant and transform covariantly under \eqref{complex coordinate transformations}. In general only the separations $\x_{12}$ have this property, however upon fixing the special configuration $\zbar_1=\zbar_2=\zbar_3\equiv \zbar$, we can also consider the quantity
	\begin{equation}
	F_{123}=u_1 z_{23}+u_2 z_{31}+u_3 z_{12}\,,
	\end{equation}
	which transforms as
	\begin{equation}
	\begin{aligned}
	F_{123}'&=e^{i\theta} F_{123}\,, &\qquad &(\text{rotation})\,,\\
	F_{123}'&=\lambda^2 F_{123}\,, &\qquad &(\text{dilation})\,,\\
	F_{123}'&=\frac{F_{123}}{1-k \zbar}\,,  &\qquad &(\text{SCT})\,,\\
	F_{123}'&=\frac{F_{123}}{(1-\bar k z_1)(1-\bar k z_2)(1-\bar k z_3)}\,, &\qquad &(\text{SCT})\,,
	\end{aligned}
	\end{equation}
	while it is invariant under all remaining symmetries. From this we are able to write down the chiral three-point functions, by demanding that correlation functions transform like the fields in \eqref{conformal transfo}. Translation and carroll boost invariance imply that the chiral 3-point function is a function of the coordinates through $z_{ij}$ and $F_{123}$, while covariance under conformal transformations fixes its form to be
	\begin{equation}
	\langle O_1 O_2 O_3 \rangle=c_{123}\, \frac{\delta(\zbar_{12})\delta(\zbar_{23})}{ (z_{12})^a\, (z_{23})^b\, (z_{13})^c\, (F_{123})^d}\,.
	\end{equation}
	Specifically, covariance under dilation and rotation requires
	\begin{align}
	\begin{split}
	a+b+c+2d+2&=\Delta_1+\Delta_2+\Delta_3\,, \\
	a+b+c+d-2&=J_1+J_2+J_3\,.
	\end{split}
	\end{align}
	Special conformal transformations generated by $k, \bar k$ respectively imply
	\begin{equation}
	\label{d parameter}
	d=2\bar h_1+2\bar h_2+2\bar h_3-4\,,
	\end{equation}
	and
	\begin{equation}
	a+c+d=2h_1\,, \qquad a+b+d=2h_2\,, \qquad b+c+d=2h_3\,.
	\end{equation}
	The unique solution to these constraints is
	\begin{equation}
	\label{abc parameters}
	a=J_1+J_2-\Delta_3+2\,, \qquad b=J_2+J_3-\Delta_1+2\,, \qquad c=J_1+J_3-\Delta_2+2\,,
	\end{equation}
	such that we can write
	\begin{equation}
	\label{generic 3-point function}
	\langle O_1 O_2 O_3 \rangle=\frac{c_{123}\, \delta(\zbar_{12})\delta(\zbar_{23})}{ (z_{12})^{J_1+J_2-\Delta_3+2}\, (z_{23})^{J_2+J_3-\Delta_1+2}\, (z_{13})^{J_1+J_3-\Delta_2+2}\, (F_{123})^{2(\bar h_1+\bar h_2+\bar h_3-2)}}\,,
	\end{equation}
	together with the complex conjugate solution. Correlators of this kind have appeared in the works~\cite{Banerjee:2019prz,Salzer:2023jqv,Mason:2023mti}.
	
	For completeness, we also present two more types of carrollian three point functions. The first one is a generalization of a three-point function with real kinematics as given in \cite{Nguyen:2023miw} to spinning operators, namely  
	\begin{equation}
	\label{eq:3pt2delta}
	\langle O_1O_2O_3\rangle=c_{123}\,\frac{\delta(z_{12})\delta(\zb_{12})\, \delta_{J_3,J_1+J_2}}{u^{\Delta_1+\Delta_2-\Delta_3-2}_{12} z^{2h_3}_{23}\zbar^{2\hb_3}_{23}}\,.
	\end{equation}
	The second one has complex kinematics and is given by
	\begin{equation}
	\label{eq:3pt3delta}
	\langle O_1O_2O_3\rangle=c_{123}\, \frac{\delta(z_{12})\delta(\zbar_{12})\delta(z_{13})\, \delta_{J_1-J_2+\Delta_3,1}}{u^{2(-2+h_1+h_2+h_3)}_{12}\zbar^{2\hb_3}_{13}}\,.
	\end{equation}
	To the best of our knowledge, this has not appeared in the literature before.

	\subsection{4-point functions}
	Momentum conservation with four momenta, when expressed in complex stereographic coordinates, amounts to \cite{Pasterski:2017ylz}
	\begin{equation}
	\label{z=zbar}
	z=\zbar\,,
	\end{equation}
	where $z, \zbar$ are the invariant cross ratios
	\begin{equation}
	z= \frac{z_{12}z_{34}}{z_{13}z_{24}}\,, \qquad \zbar= \frac{\zbar_{12}\zbar_{34}}{\zbar_{13}\zbar_{24}}\,.
	\end{equation}
	In the context of scattering amplitudes we are thus interested in 4-point functions featuring a Dirac distribution $\delta(z-\zbar)$. For later use we also recall the useful relations
	\begin{equation}
	1-z=\frac{z_{14}z_{23}}{z_{13}z_{24}}\,, \qquad \frac{1-z}{z}=\frac{z_{14}z_{23}}{z_{12}z_{34}}\,.
	\end{equation}
	We again look for combinations of the coordinates $\x_i$ which are translation-invariant and transform covariantly under \eqref{complex coordinate transformations}. In addition to the separations $\x_{ij}$, on the support of \eqref{z=zbar} we also have the interesting combination
	\begin{equation}
	\label{F1234}
	\begin{split}
	F_{1234}&\equiv u_4-u_1 z \left| \frac{z_{24}}{z_{12}}\right|^2+u_2 \frac{1-z}{z}\left| \frac{z_{34}}{z_{23}}\right|^2-u_3 \frac{1}{1-z}\left| \frac{z_{14}}{z_{13}}\right|^2\\
	&=u_4-u_1\, \frac{z_{34} \zbar_{24}}{z_{13}\zbar_{12}}+u_2\, \frac{z_{14} \zbar_{34}}{z_{12}\zbar_{23}}-u_3\, \frac{z_{24}\zbar_{14}}{z_{23}\zbar_{13}}\,.
	\end{split}
	\end{equation}
	Note that, on the support \eqref{z=zbar}, any permutation on the indices yields a quantity related to \eqref{F1234} by a simple multiplicative factor, for example
	\begin{equation}
	F_{4231}=-\frac{1}{z} \left|\frac{z_{12}}{z_{24}} \right|^2 F_{1234}\,, \qquad F_{2143}=-\frac{z_{13}\zbar_{23}}{z_{14}\zbar_{24}} F_{1234}\,, \qquad F_{1432}=\frac{z_{12}\zbar_{23}}{z_{14}\zbar_{34}} F_{1234}\,,
	\end{equation}
	where we note that under $1 \leftrightarrow 4$ we also have $z \leftrightarrow z^{-1}$.
	Therefore we can restrict our attention to $F_{1234}$ without loss of generality, which can be shown to follow the simple transformation rules
	\begin{equation}
	\begin{aligned}
	F_{1234}'&=\lambda F_{1234}\,, &\qquad &(\text{dilation})\,,\\
	F_{1234}'&=\frac{F_{1234}}{1-k \zbar_4}\,, \quad F_{1234}'=\frac{F_{1234}}{1-\bar k z_4}\,, &\qquad &(\text{SCT})\,.
	\end{aligned}
	\end{equation}
	On the support $z=\zbar$, they are invariant under all other transformations \eqref{complex coordinate transformations}. 
	
	Thus we can assume an ansatz of the form
	\begin{equation}
	\label{4-point general}
	\langle O_1 O_2 O_3 O_4 \rangle= \delta(z-\zbar) G(z)\prod_{i<j} \frac{1}{(z_{ij})^{a_{ij}} (\zbar_{ij})^{\bar{a}_{ij}}(F_{1234})^c}\,,
	\end{equation}
	with $G(z)$ an arbitrary function of the invariant cross ratio, as required by translation and carroll boost invariance. Enforcing covariance under dilation and rotation yields the constraints 
	\begin{equation}
	\sum_{i<j} (a_{ij}+\bar{a}_{ij})+c=\sum_i \Delta_i\,, \qquad \sum_{i<j} (a_{ij}-\bar{a}_{ij})=\sum_i J_i\,.
	\end{equation}
	Requiring covariance under SCT yields
	\begin{equation}
	\sum_{i\neq j} a_{ij}=2h_j \quad (j\neq 4)\,, \qquad 
	\sum_{i \neq 4}a_{i4}+c=2h_4\,,
	\end{equation}
	together with the conjugate relations. 
	The solution to these constraints is given by
	\begin{align}
	\label{aij}
	\begin{split}
	a_{ij}&=h_i+h_j-H/3+c/6\,, \quad (i,j \neq 4)\,,\\
	a_{i4}&=h_i+h_4-H/3-c/3\,,
	\end{split}
	\end{align}
	with $H\equiv \sum_i h_i$, and conjugate relations. Note that we are left with one free parameter $c$. If $c=0$ then \eqref{4-point general} reduces to a standard chiral four-point function of a $\operatorname{CFT}_2$.
	
	\section{Carrollian amplitudes}
	\label{section 4}
	The modified Mellin transform \eqref{Mellin} can be applied to momentum $\mathcal{S}$-matrix elements $S_n$, thereby defining the \textit{carrollian amplitudes}
	\begin{equation}
	\label{Mellin amplitudes}
	\langle O^{\eta_1}_{\Delta_1,J_1}(x_1^\alpha)\, ...\, O^{\eta_n}_{\Delta_n,J_n}(x_n^\alpha) \rangle\equiv \prod_{k=1}^n \int_0^\infty d\omega_k\, \omega^{\Delta_k-1} e^{i \eta_k \omega_k u_k} S_n(1^{J_1}...\,n^{J_n})\,,
	\end{equation}
	where $\eta_k=\pm 1$ depending whether the particle is ingoing (+) or outgoing (-), with momenta parametrised as 
	\begin{equation}
	p_k^\mu=\eta_k\, \frac{\omega_k}{\sqrt{2}}(1+x_k^2,2x_k^i,1-x_k^2)\,.
	\end{equation}
	This is a convention where all particles can be effectively treated as if they were ingoing, with ingoing momenta $p_k^\mu$ as given above and ingoing helicity $J_k$. We emphasise that the in/out label $\eta$ distinguishes operators of distinct `flavor', although we will often drop it for notational convenience.
	Simply based on the transformation properties of the $\mathcal{S}$-matrix elements, the carrollian amplitudes necessarily transform as correlation functions for the corresponding carrollian conformal fields. In this section we apply \eqref{Mellin amplitudes} to a variety of 2-, 3- and 4-point scattering amplitudes of massless particles, following earlier works \cite{Banerjee:2019prz,Bagchi:2022emh,Donnay:2022wvx,Mason:2023mti,Bagchi:2023cen,Bagchi:2023fbj}. We show that they provide examples of the general correlation functions constructed in section~\ref{section 3}.
	
	\subsection{2-point amplitudes} 
	We start with the two-point carrollian amplitude, which is the modified Mellin transform of the 1-1 scattering amplitude, with $\eta_2=1=-\eta_1$, equal to the Lorentz-invariant inner product,
	\begin{equation}
	S_2(1^{J_1}2^{J_2})=|\vec p_1| \delta(\vec p_1+\vec p_2) \delta_{J_1,-J_2}=\omega_1{}^{-1} \delta(\omega_1-\omega_2) \delta(x_1^i-x_2^i)\delta_{J_1,-J_2}\,,
	\end{equation}
	where the last expression follows from the momentum parametrisation \eqref{momentum parametrisation}.
	Application of \eqref{Mellin amplitudes} yields \cite{Banerjee:2018gce,Bagchi:2022emh,Liu:2022mne,Donnay:2022wvx}
	\begin{equation}
	\langle O_1 O_2 \rangle= \Gamma[\Delta_1+\Delta_2-2] \frac{\delta(x_{12}^i)\delta_{J_1,-J_2}}{(iu_{12})^{\Delta_1+\Delta_2-2}}\,,     
	\end{equation}
	which is manifestly of the general form \eqref{2-point function 1}. Note that the above expression diverges for $\Delta_1+\Delta_2=2$ due to the pole in the Gamma function. This divergence can be matched to an anomalous $\ln r$ divergence in the carrollian two-point function obtained from the extrapolate holographic dictionary \cite{Nguyen:2023miw}. For a well-defined two-point function we should therefore consider $\Delta_1+\Delta_2\neq 2$.
	
	\subsection{3-point amplitudes} 
	As discussed at the beginning of section~\ref{subsection 3.2}, momentum conservation for three massless particles only leaves us with amplitudes that are rather singular if only real kinematics are considered. With complex kinematics there exist 3-point amplitudes which are regular in the sense that they do not contain additional delta functions apart from the usual one enforcing momentum conservation. Even though they may appear unphysical, they constitute important building blocks to construct higher-point amplitudes through recursive equations. Furthermore their form is entirely fixed by the little group scalings and locality of the interaction, which is most conveniently displayed in spinor-helicity variables \cite{Elvang:2015rqa,Badger:2023eqz}
	\begin{align}
	\label{spinhel3pt}
	S_3(1^{J_1}2^{J_2}3^{J_3})=\!\!\begin{cases} \braket{12}^{J_3-J_1-J_2}\braket{31}^{J_2-J_1-J_3}\braket{23}^{J_1-J_2-J_3}\delta(\Sigma_k\, p_k),\!\!\! &J_1+J_2+J_3<0\,,\\
	[12]^{-J_3+J_1+J_2}[31]^{-J_2+J_1+J_3}[23]^{-J_1+J_2+J_3}\, \delta(\Sigma_k\, p_k),\!\!\! &J_1+J_2+J_3>0\,,\end{cases}
	\end{align}
	up to an overall free coefficient. As shown in \cite{Pasterski:2017ylz} the spinor-helicity variables can be chosen such that $\langle ij\rangle=\sqrt{\omega_i\omega_j}\,z_{ij}$ and $[ij]=-\eta_i\eta_j\sqrt{\omega_i\omega_j}\, \zb_{ij}$. The modified Mellin transform of \eqref{spinhel3pt} has been performed with $\Delta_k=1$ in \cite{Salzer:2023jqv,Mason:2023mti}. Generalising their computation to arbitrary scaling dimensions yields, for $J_1+J_2+J_3<0$, 
	\begin{align}
	\label{eq:3ptwithTheta}
	\begin{split}
	\langle O_1 O_2 O_3 \rangle &=\Gamma[2\Sigma_k \bar h_k-4]\, \Theta\left(-\frac{z_{13}}{z_{23}}\eta_1\eta_2\right)\Theta\left(\frac{z_{12}}{z_{23}}\eta_1\eta_3\right)\\
	&\times \frac{\delta(\zbar_{12}) \delta(\zbar_{13}) (z_{12})^{\Delta_3-J_1-J_2-2} (z_{23})^{\Delta_1-J_2-J_3-2} (z_{13})^{\Delta_2-J_1-J_3-2}}{\left(z_{23}\, u_1-z_{13}\, u_2+z_{12}\, u_3 \right)^{2\Sigma_k \bar h_k-4}}\,,
	\end{split}
	\end{align}
	again up to an overall constant coefficient. We see that this is indeed of the general form \eqref{generic 3-point function} derived in the previous section. The expression for $J_1+J_2+J_3>0$ is obtained by the replacement $z_k \to \zbar_k$ and $h_k \to \bar h_k$.
	
	\subsection{4-point tree-level amplitudes}
	We now look at some important examples of 4-point tree-level amplitudes, namely the scalar contact amplitude, and the gluon and graviton MHV amplitudes. The computation of their modified Mellin transform will closely follow the methodology of used in \cite{Mason:2023mti}. In particular writing the $\mathcal{S}$-matrix element as $S_4=A_4\,  \delta(\Sigma_k p_k)$ and using the following representation of the momentum-conserving delta function,
	\begin{align}
	\label{delta p4}
	\begin{split}
	\delta(\Sigma_k p_k)&=\frac{\delta(z-\zbar)}{4\omega_4|z_{13}z_{24}|^2}\, \delta\left(\omega_1+z\left|\frac{z_{24}}{z_{12}}\right|^2 \eta_1 \eta_4 \omega_4 \right)\\
	&\times\delta\left(\omega_2-\frac{1-z}{z} \left|\frac{z_{34}}{z_{23}}\right|^2\eta_2 \eta_4 \omega_4 \right)\delta\left(\omega_3+\frac{1}{1-z}\left|\frac{z_{14}}{z_{13}}\right|^2\eta_3 \eta_4 \omega_4  \right)\,,
	\end{split}
	\end{align}
	application of \eqref{Mellin amplitudes} directly yields, up to a constant phase,
	\begin{align}
	\label{C4 general}
	\begin{split}
	C_4&=\delta(z-\zbar)\,\Theta\left(-z\left|\frac{z_{24}}{z_{12}}\right|^2 \eta_1 \eta_4  \right)\Theta\left(\frac{1-z}{z} \left|\frac{z_{34}}{z_{23}}\right|^2\eta_2 \eta_4  \right)\Theta\left(-\frac{1}{1-z}\left|\frac{z_{14}}{z_{13}}\right|^2\eta_3 \eta_4 \right)\\
	&\times \frac{z^{\Delta_1-\Delta_2} (1-z)^{\Delta_2-\Delta_3}}{|z_{13}z_{24}|^2} \left|\frac{z_{24}}{z_{12}} \right|^{2(\Delta_1-1)}\left|\frac{z_{34}}{z_{23}} \right|^{2(\Delta_2-1)} \left|\frac{z_{14}}{z_{13}} \right|^{2(\Delta_3-1)}\\
	&\times\int_0^\infty d\omega_4\, \omega_4^{\Sigma \Delta-5} e^{i\eta_4\omega_4 F_{1234}} A_4^*\,,
	\end{split}
	\end{align}
	where $F_{1234}$ is the quantity defined in \eqref{F1234}, and $A_4^*$ is the scattering amplitude evaluated on the support of \eqref{delta p4}. Provided $A_4^*$ is a polynomial in $\omega_4$, the remaining integral can be evaluated using the formula
	\begin{equation}
	\label{Schwinger}
	\int_0^\infty d\omega\, \omega^{\Delta-1}\, e^{-i\omega u}=\frac{\Gamma[\Delta]}{(iu)^\Delta}\,, \qquad \text{Im}(u)<0\,.
	\end{equation}
	
	\paragraph{Scalar contact amplitude.} The simplest example of 4-particle scattering amplitude one can think of is the contact amplitude corresponding to $\lambda \phi^4$ interaction, given by $A_4=\lambda$. Plugging this into \eqref{C4 general} and using \eqref{Schwinger} we directly obtain the carrollian amplitudes
	\begin{align}
	\begin{split}
	C_4&=\delta(z-\zbar)\Theta\left(-z\left|\frac{z_{24}}{z_{12}}\right|^2 \eta_1 \eta_4  \right)\Theta\left(\frac{1-z}{z} \left|\frac{z_{34}}{z_{23}}\right|^2\eta_2 \eta_4  \right)\Theta\left(-\frac{1}{1-z}\left|\frac{z_{14}}{z_{13}}\right|^2\eta_3 \eta_4 \right)\\
	&\times \frac{z^{\Delta_1-\Delta_2} (1-z)^{\Delta_2-\Delta_3}}{|z_{13}z_{24}|^2} \left|\frac{z_{24}}{z_{12}} \right|^{2(\Delta_1-1)}\left|\frac{z_{34}}{z_{23}} \right|^{2(\Delta_2-1)} \left|\frac{z_{14}}{z_{13}} \right|^{2(\Delta_3-1)}\frac{\Gamma[\Sigma \Delta-4]}{(iF_{1234})^{\Sigma \Delta-4}}\,.
	\end{split}
	\end{align}
	Although this formula looks relatively cumbersome at first sight, we can equivalently write it in general form \eqref{4-point general} derived in section~\ref{section 3}, 
	\begin{equation}
	\label{C4 scalar}
	C_4=  \Gamma[c]\, \Theta\left(...  \right)\Theta\left(...  \right)\Theta\left(... \right) \prod_{i<j} \frac{G(z)\delta(z-\zbar)}{(z_{ij})^{a_{ij}} (\zbar_{ij})^{\bar{a}_{ij}}(iF_{1234})^c}\,,
	\end{equation}
	with 
	\begin{equation}
	G(z)=\left[z(1-z)\right]^{2/3}\,, \qquad c=\Sigma \Delta-4\,,
	\end{equation}
	and all other parameters $a_{ij}$ determined as in \eqref{aij}. This demonstrates the usefulness of \eqref{4-point general} in organising the possible four-point carrollian amplitudes.
	
	\paragraph{Gluon and graviton MHV amplitudes.} The four-point (color-ordered) gluon and graviton MHV amplitude are given by \cite{Elvang:2015rqa}
	\begin{align}
	A^{\text{YM}}_4(1^{+1}2^{-1}3^{-1}4^{+1})&=\frac{\langle 23 \rangle^4}{\langle 12 \rangle \langle 23\rangle \langle 34\rangle \langle 41\rangle}=\frac{\omega_2\omega_3}{\omega_1\omega_4} \frac{z_{23}^3}{z_{12}z_{34}z_{41}}\,,\\
	A_4^{\text{GR}}(1^{+2}2^{-2}3^{-2}4^{+2})&=\frac{\langle 23 \rangle^7 [23]}{\langle 13\rangle \langle 34\rangle \langle 12 \rangle \langle 24 \rangle \langle 14 \rangle^2}=\frac{(\omega_2 \omega_3)^3}{(\omega_1\omega_4)^2}\frac{z_{23}^7 \zbar_{23}}{z_{13}z_{34}z_{12}z_{24}z_{14}^2}\,.
	\end{align}
	Applying \eqref{C4 general}-\eqref{Schwinger}, the carrollian amplitudes we obtain are of the form \eqref{C4 scalar} with 
	\begin{align}
	G^{\text{YM}}_{+--+}(z)&=z^{-1/3} (1-z)^{5/3}\,, &c^{\text{YM}}=\Sigma \Delta-4\,,\\
	G^{\text{GR}}_{+--+}(z)&=z^{-2/3} (1-z)^{10/3}\,,   &c^{\text{GR}}=\Sigma \Delta-2\,.
	\end{align}
	The other helicity configurations can be obtained by renaming the indices. This can be done easily by noticing that the denominator in the general formula \eqref{4-point general} carries all the kinematic structure. Hence the unconstrained combination $\delta(z-\zbar) G(z)$ alone induces a non-trivial change of expression under such renaming of indices. Under $2 \leftrightarrow 4$ the cross ratio transforms as $z \leftrightarrow 1-z$ such that we obtain
	\begin{align}
	G^{\text{YM}}_{++--}(z)&=z^{5/3} (1-z)^{-1/3}\,,\\
	G^{\text{GR}}_{++--}(z)&=z^{10/3} (1-z)^{-2/3}\,.
	\end{align}
	Under $3 \leftrightarrow 4$ the cross ratio transforms as $z \leftrightarrow z/(z-1)$ with $\delta(z-\zbar) \leftrightarrow (1-z)^2 \delta(z-\zbar)$, such that we obtain
	\begin{align}
	G^{\text{YM}}_{+-+-}(z)&=(-z)^{-1/3} (1-z)^{2/3}\,,\\
	G^{\text{GR}}_{+-+-}(z)&=(-z)^{-2/3} (1-z)^{-2/3}\,.
	\end{align}
	This provides explicit examples of four-point carrollian correlators of the general form \eqref{4-point general}.
	
	\section{Carrollian OPE structures}
	\label{section 5}
	While in the previous sections we mostly discussed kinematic constraints on carrollian correlators, it is now time to address the structure of interactions and the constraints they impose on the spectrum of operators. 
	
	One of the pillars of standard conformal field theory is the operator product expansion (OPE), which allows to express the product of two local operators as a sum of local operators,
	\begin{equation}
	O_1(\vec x_1)\, O_2(\vec x_2)= \sum_{k} C_{12k}(\vec x_{12})\, O_k(\vec x_2) \,,
	\end{equation}
	where the sum is over primary operators and descendant operators.  
	This equality is made possible by the state-operator correspondence which expresses the fact that any quantum state can be created from insertion of a local operator at the point $\vec x_2$. In the coincidence limit $\vec x_{12} \to 0$, the OPE takes the simple form
	\begin{equation}
	\label{standard OPE coincident}
	O_1(\vec x)\, O_2(0)  \stackrel{\vec x \sim 0}{\approx } \sum_{k} \frac{c_{12k}}{|\vec x|^{\Delta_1+\Delta_2-\Delta_k}}\, O_k(0)+ subleading\,,
	\end{equation}
	where the subleading terms contain derivatives of the primary operators and therefore account for their descendants. The latter are actually completely fixed by conformal symmetry, such that the set of coefficients $\{c_{12k}\}$ carry all the independent data. 
	
	In this work we wish to investigate the existence of an analogous structure within carrollian conformal field theory. For simplicity we will first focus our attention on the weaker form of the OPE, i.e., that arising in a coincidence limit of the kind \eqref{standard OPE coincident}. But first we need to specify what we mean by `coincidence limit' in a carrollian setting. Given a product of two local operators $O_1(\x_1)\, O_2(\x_2)$, we will in fact consider two kinds of limits :
	\begin{itemize}
		\item[1.] The \textit{uniform} coincidence limit $\x_{12} \to 0$ where the operators truly collide.
		\item[2.] The \textit{holomorphic} coincidence limit $z_{12} \to 0$ with finite separations $\zbar_{12} \neq 0$ and $u_{12} \neq 0$, and the analogous anti-holomorphic coincidence limit. 
	\end{itemize}
	Both situations correspond to vanishing of the invariant distance between the two operators insertions, as can be easily seen from the metric \eqref{scri metric}. 
	
	We will systematically construct the leading terms of a consistent OPE for the uniform coincidence limit, whose study was already initiated in \cite{Banerjee:2020kaa}. We will uncover a significantly more complex structure than in the standard case \eqref{standard OPE coincident}. One important complication comes from the fact that a primary operator $O_k$ of dimension $(h_k,\bar h_k)$ may \textit{descend} from another primary operator $O_{k'}$ of dimension $(h_{k'},\bar h_{k'})=(h_k-n/2,\bar h_k-n/2)$ if they satisfy $O_k=(\partial_u)^n\, O_{k'}$. Hence there is a priori no absolute primary within a carrollian conformal block. The second source of complexity comes from the fact that, as with correlation functions, the form of the leading term in the OPE is not completely fixed by symmetry. This leads to various possible OPE branches for a fixed $O_k$. Of course knowledge of the 3-point function $\langle O_1 O_2 O_k \rangle$ would determine the leading OPE coefficient and would thus select a particular OPE branch.
	
	While the uniform coincidence limit is perhaps the most natural one to study, the holomorphic coincidence limit has recently been discussed in relation to the colinear factorisation of tree-level massless scattering amplitudes \cite{Mason:2023mti}. Specifically, starting from the well-knwon colinear factorisation of momentum space amplitudes, the authors of \cite{Mason:2023mti} derived a specific form of holomorphic carrollian OPE satisfied by carrollian amplitudes. Using symmetry alone, here we will construct a holomorphic OPE which contains the one presented in \cite{Mason:2023mti} as a particular case, before discussing its extension to subleading orders in $z_{12} \sim 0$.
	
	Finally we will discuss the form of the carrollian OPE blocks for finite separation $\x_{12} \neq 0$, adapting the construction in \cite{Czech:2016xec}. The resulting carrollian OPE blocks will be compatible with the \textit{celestial} OPE blocks discussed in \cite{Guevara:2021tvr}. Although this is not an easy task, we will look at the uniform coincidence limit of these OPE blocks and in some cases recover results established in previous sections.  
	
	\subsection{Uniform coincidence limit}
	\label{section 5.1}
	In analogy with \eqref{standard OPE coincident}, we postulate the existence of an OPE of the form
	\begin{equation}
	\label{Carrollian OPE}
	O_1(\x)\, O_2(0) \stackrel{\x\sim 0}{\approx} \sum_k f_{12k}(\x)\, O_k(0)+subleading+massive\,,
	\end{equation}
	where the sum is over \textit{single-particle} carrollian  primary fields. As usual the subleading terms involve the descendants operators \eqref{descendant fields}. Unlike in conventional CFT where one can rely on the state-operator correspondence, in this context it is a priori unclear whether there exists a convergent OPE and what is the full set of operators which need to be considered on the right-hand side of \eqref{Carrollian OPE}. In particular, the `massive' terms may correspond to at least two different types of operators. First they can correspond to massive one-particle operators, for instance in the context of a scattering theory involving massive particles, in which case they cannot be local carrollian operators of the type considered in this paper.\footnote{They are local carrollian operators on $\mathsf{Ti}$ \cite{Have:2024dff}.} Second they may correspond to multi-particle states. Although we will not consider the corresponding OPE blocks explicitly in this work, at the end of this subsection we discuss their unavoidable appearance. Regardless we directly proceed to constrain the functions $f_{12k}(\x)$ by requiring consistency with Poincar\'e symmetry. In practice we act on both sides of \eqref{Carrollian OPE} with the symmetry generators and require consistency order by order in $\x \sim 0$. 
	
	\subsubsection*{Several OPE branches} 
	We determine the explicit form of $f_{123}$ allowed by symmetry, focusing on the contribution from a single primary operator $O_3$. Acting with the generators $\{ H,P_i\}$ does not yield any constraint since our ansatz already incorporates carrollian translation invariance. Acting with the generators $\{K,K_i,B_i\}$ on either side of \eqref{Carrollian OPE} does not contribute at leading order in $x^i \sim 0$ as can be seen from \eqref{Carrollian induced rep}. Therefore we are left to act with $\{ D,J_{12} \}$ or equivalently with $\{L_0,\bar L_0\}$. Acting with $L_0$ on the left and on the right of \eqref{Carrollian OPE} yields, respectively,
	\begin{equation}
	\begin{split}
	\left[L_0,O_1(\x)\, O_2(0) \right]&=\left(\frac{u}{2}\partial_u+z\partial_z+h_1+h_2 \right)O_1(\x)\, O_2(0)\\
	&\approx \left(\frac{u}{2}\partial_u+z\partial_z+h_1+h_2 \right) f_{123}(\x)\, O_3(0)\,,
	\end{split}
	\end{equation}
	and
	\begin{equation}
	f_{123}(\x) \left[L_0,O_3(0) \right]=h_3\, f_{123}(\x)\, O_3(0)\,.
	\end{equation}
	Hence we should impose
	\begin{equation}
	\label{h definition}
	\left(\frac{u}{2}\partial_u+z\partial_z-h \right) f_{123}(\x)=0\,, \qquad h\equiv h_3-h_1-h_2\,,
	\end{equation}
	which essentially tells us that $f_{123}$ must have scaling weight $h=h_3-h_1-h_2$ under holomorphic scalings generated by $L_0$ (and similarly for $\bar L_0$). The general form satisfying this property is
	\begin{equation}
	\label{f12k}
	\begin{split}
	f_{123}(\x)&=c_0\, z^{h-a} z^{\bar h-a} u^{2a}+c_1\, \delta(z) \delta(\zbar)\, u^{h+\bar h+2}\\
	&+c_2\, \delta(\zbar) z^{h-\bar h-1} u^{2\bar h+2}+\bar c_2\, \delta(z) \zbar^{\,\bar h- h-1} u^{2h+2}\,,
	\end{split}
	\end{equation}
	where the coefficients $c_0,c_1,c_2,\bar c_2$ as well as the exponent $a$ are arbitrary numbers.

	\subsubsection*{Parent and ancestor primaries}
	Given a primary operator $O_3$ appearing on the right-hand side of \eqref{Carrollian OPE} with one of the allowed leading OPE functions \eqref{f12k}, we can start studying the operators appearing at subleading orders in the expansion variables $z,\zbar,u$. In standard conformal field theory, there is a finite number of operators which can appear at a given order. This is not the case anymore, since the operator $O_3$ of dimension $(h_3,\bar h_3)$ may possess \textit{parent primary operators} $O_{3'}$ of dimension $(h_{3'},\bar h_{3'})=(h_3-n/2,\bar h_3-n/2)$ in case they satisfy $O_3=(\partial_u)^n\, O_{3'}$. If one allows for all `ancestors' without further restriction, then at a given order in the OPE expansion there are in principle infinitely many descendants which may appear. As we do not wish to tackle a problem of infinite complexity, we will consider the simplest nontrivial case where only the first parent $O_{3'}$ satisfying $O_3=\partial_u O_{3'}$ is allowed to enter the game. This was already considered in the analysis presented in \cite{Banerjee:2020kaa}, which we will extend.
	
	Given the two primary operators $O_3$ and $O_{3'}$ related by $\partial_u O_3'=O_3$, we want to list the descendant operators that may appear at a given order in the OPE expansion. Following \cite{Banerjee:2020kaa}, we will consider all BMS descendants \eqref{descendant fields} rather than just the Poincar\'e descendants. Although this might be surprising, we will see that the supertranslation descendants are absolutely necessary except in very fine-tuned situations. Of course the conformal group of $\scri$ being the full BMS group, it is also sensible to introduce them as part of the carrollian CFT construction. 
	
	If an operator $O_3$ has weights $(h_3,\hb_3)$, then by acting with the BMS generators we obtain descendant operators with weights
	\begin{equation}
	\begin{split}
	L_nO_3\qquad &(h_3-n\,,\hb_3)\,,\\
	P_{m,n}O_3\qquad &(h_3-m-1/2\,,\hb_3-n-1/2). 
	\end{split}
	\end{equation}
	When evaluated at $\x=0$, due to the appearance of positive powers of $z,\zbar$ in \eqref{eq:supertrans}, we have 
	\begin{equation}
	L_n O_3(0)=0\,, \qquad P_{m,n}O_3(0)=0\,, \qquad m\geq 0 \lor n \geq 0\,.
	\end{equation}
	On the other hand the operators $L_nO_3(0)$ and $P_{m,n}O_3(0)$ would appear ill-defined for $m\leq -2 \lor n \leq -2$ due to the appearance of negative powers $z, \zbar$ in \eqref{eq:supertrans}. As done in \cite{Banerjee:2020kaa}, one should therefore only consider these operators when inserted inside correlation functions, with 
	\begin{equation}
	\label{supertranslation Ward id}
	\langle P_{m,n} O_3(0) \prod_{i=1}^N O_i(\x_i) \rangle=\sum_{j=1}^N z_j^{m+1} \zbar_j^{n+1} \partial_{u_j} \langle O_3(0) \prod_{i=1}^N O_i(\x_i) \rangle\,, \quad (m\leq -2 \lor n \leq -2)\,,
	\end{equation}
	which can be recognized as the supertranslation Ward identity. There is a technicality worth mentioning at this point. In the correlation functions appearing on the right-hand side of \eqref{supertranslation Ward id}, we generically expect terms containing Dirac distributions $\delta(z_j)$, such that it is primordial to specify the distributional meaning of $z_j^{m+1}$ when $m\leq -2$. The only way that these distributional products are well-defined is if the singularity is removed, namely by defining it to be the pseudo-function \cite{Kanwal,Nguyen:2023miw}
	\begin{equation}
	\frac{1}{z^n}\equiv \text{Pf}\left(\frac{1}{z^n} \right)\,,
	\end{equation}
	which in particular coincides with Cauchy's principal value for $n=1$. This yields the simple distributional equality
	\begin{equation}
	\label{zn delta}
	z^{-n} \delta(z)=0 \qquad (n\neq 0)\,.
	\end{equation}
	
	Let us now list the operators descending from $O_3$ and $O_{3'}$ that can appear at the first subleading orders, i.e., with scaling dimension between $\Delta_3$ and $\Delta_3+2$. We find 
	\begin{equation}
	\begin{split}
	(h_3+1,\hb_3):& \quad L_{-1}O_3\,, P_{-2,-1}O_{3'}\\
	(h_3,\hb_3+1):& \quad \bar{L}_{-1}O_3\,, P_{-1,-2}O_{3'}\\
	(h_3+\frac{1}{2},\hb_3+\frac{1}{2}):& \quad P_{-1,-1}O_3\,,L_{-1}\bar L_{-1} O_{3'}\\
	(h_3+1,\hb_3+1):&\quad L_{-1}\bar{L}_{-1}O_3\,,P^2_{-1,-1}O_3\,,P_{-2,-2}O_{3'}\,,L_{-1}P_{-1,-2}O_{3'}\,,\bar{L}_{-1}P_{-2,-1}O_{3'}\,\\
	(h_3+\frac{3}{2},\hb_3+\frac{1}{2}):&\quad P_{-2,-1}O_3\,, L_{-1}P_{-1,-1}O_3\,, L^2_{-1}\bar{L}_{-1}O_{3'}\,, L_{-2}\bar{L}_{-1} O_{3'}\\
	(h_3+\frac{1}{2},\hb_3+\frac{3}{2}):&\quad P_{-1,-2}O_3\,, \bar{L}_{-1}P_{-1,-1}O_3\,, \bar{L}^2_{-1}L_{-1}O_{3'}\,, \bar{L}_{-2}L_{-1} O_{3'}
	\end{split}
	\end{equation}
	Armed with this list of operators, we can look at the subleading terms in the OPE of any one of the branches corresponding to the coefficients $c_0, c_1, c_2,\bar c_2$ in \eqref{f12k}. 
	
	\subsubsection*{Regular OPE}
	As an important case, let us first study the OPE branch with $c_0 \neq 0$. Using the above list of operators, we write the ansatz 
	\begin{align}
	&O_1(\x)O_2(0)\sim z^{h-a}\, \zb^{\hb-a}\, u^{2a}\Big[O_3+u ( \beta_1 P_{-1,-1}O_3+\beta_2 L_{-1}\bar L_{-1} O_{3'})\nonumber\\
	\label{OPE ansatz}
	&+z(\alpha_1 L_{-1}O_3+\alpha_2P_{-2,-1}O_{3'})+\bar{z}(\bar{\alpha}_1 \bar{L}_{-1}O_3+\bar{\alpha}_2P_{-1,-2}O_{3'})
	\\
	&+ z\zbar (\gamma_1L_{-1}\bar{L}_{-1}O_3
	+\gamma_3 P^2_{-1,-1}O_3+\gamma_2P_{-2,-2}O_{3'}+\gamma_4 L_{-1}P_{-1,-2}O_{3'}+\bar{\gamma}_4 \bar{L}_{-1}P_{-2,-1}O_{3'})\nonumber\\
	&+...\,\Big](0)\,, \nonumber
	\end{align}
	where all operators on the right-hand side are evaluated at the origin, and where $h, \bar h$ are defined as in \eqref{h definition}.
	Acting with $P_{-1,0}\,,P_{0,-1}\,,P_{0,0}$ and imposing consistency of the expansion \eqref{OPE ansatz} results in the conditions\footnote{We note that this differs from the result presented in \cite{Banerjee:2020kaa}, where consistency with the action of $P_{0,-1}$ is claimed to fix $\alpha_1$ in terms of the normalisation of $O_3$, while we find that it rather implies $\alpha_1=\beta_1$. However their analysis crucially does not include $\beta_1$ (nor any of the $\gamma_i$'s).}
	\begin{align}
	\label{OPE coefficients 1}
	a=0\,, \qquad \alpha_1=\bar{\alpha}_1=\beta_1=\gamma_1\,, \qquad \beta_2=0\,.
	\end{align}
	Acting with $L_1$ we find the conditions
	\begin{align}
	2h'\, O_3&=\alpha_1 L_1L_{-1}O_3+\alpha_2 L_1 P_{-2,-1}O_{3'}\,,\\
	2h'\, (\bar{\alpha}_1\bar{L}_{-1}O_3+\bar{\alpha}_2P_{-1,-2}O_{3'})&=\big(\gamma_1 L_1 L_{-1} \bar{L}_{-1}O_3+\gamma_2 L_1 P_{-2,-2}O_{3'}+\gamma_3 L_1 P^2_{-1,-1}O_3 \nonumber\\
	&\quad +\gamma_4 L_1L_{-1}P_{-1,-2}O_{3'}+\bar{\gamma}_4 L_1\bar{L}_{-1}P_{-2,-1}O_{3'}\big)\,,
	\end{align}
	with $h'\equiv h_1+h/2=(h_1-h_2+h_3)/2$. After using the algebra relations \eqref{[L,P]}-\eqref{[L,L]}, they yield the constraints
	\begin{equation}
	h'=\alpha_1 h_3+\alpha_2\,, \qquad h'\bar{\alpha}_1=\gamma_1 h_3+ \bar{\gamma}_4\,, \qquad  h'\bar{\alpha}_2=\gamma_2+h_3\gamma_4\,.
	\end{equation}
	Similarly acting with $\bar L_1$ yields
	\begin{equation}
	\bar h'=\bar \alpha_1 \bar h_3+\bar \alpha_2\,, \qquad \bar h'\alpha_1=\gamma_1 \bar h_3+ \gamma_4\,, \qquad \bar h'\alpha_2=\gamma_2+\bar h_3\bar \gamma_4\,.
	\end{equation}
	The solution to this set of equations is given by
	\begin{equation}
	\label{OPE coefficients 2}
	\begin{split}
	\alpha_2&=h'-  h_3 \beta_1\,,\\
	\bar \alpha_2&=\bar h'-  \bar h_3 \beta_1\,,\\
	\gamma_2&=h' \bar h'+(h_3 \bar h_3-\bar h' h_3-h'\bar h_3 ) \beta_1\,,\\
	\gamma_4&=(\bar h'-\bar h_3) \beta_1\,,\\
	\bar \gamma_4&=(h'-h_3) \beta_1\,,
	\end{split}
	\end{equation}
	with $\beta_1$ and $\gamma_3$ still undetermined. Hence we end up with some indeterminacy compared to the case of standard CFT. Let us note that all the coefficients \eqref{OPE coefficients 2} are associated with the appearance of $O_{3'}$ descendants. These do not vanish in general, except in the two fine-tuned cases where $(\beta_1,h',\bar h')=(1,h_3,\bar h_3)$ or $\beta_1=h'=\bar h'=0$. We conclude that it is generically not enough to consider descendants of $O_3$ alone. Parents typically get involved.
	
	We can complete the OPE in \eqref{OPE ansatz} with $a=0$ to all orders by using the ansatz 
	\begin{equation}
	\label{full OPE}
	\begin{split}
	O_1(\x)O_2(0)\sim z^{h}\zbar^{\hb}\sum^\infty_{k,\kb=1}&\sum^\infty_{m,n,\nb=0} \frac{\alpha^{k,\kb}_{m,n,\nb}}{m!n!\nb!}\, u^{m} z^{n+k-1}\zb^{\nb+\kb-1}\\
	&\times (P_{-1,-1})^m(L_{-1})^n(\bar{L}_{-1})^{\nb} P_{-k,-\kb} O_{3'}(0)\,,
	\end{split}
	\end{equation}
	where we introduced again as above $\partial_u O_{3'}=P_{-1,-1}O_{3'}=O_3$ so that the leading coefficient $\alpha^{1,1}_{0,0,0}$ corresponds to the leading coefficient $c_0$ in \eqref{f12k}. Note however that this is not the most general OPE, as we can see that it corresponds in particular to a situation where $\gamma_3=0$ in \eqref{OPE ansatz}.
	Acting with the symmetry generators on both sides produces recursion relations among the coefficients. In particular, invariance under $P_{-1,-1},P_{-1,0},P_{0,-1},P_{0,0}$ imposes the very restricting conditions
	\begin{equation}
	\label{eq:transinvariance}
	\alpha^{k,\kb}_{m+1,n,\nb}=\alpha^{k,\kb}_{m,n+1,\nb}=\alpha^{k,\kb}_{m,n,\nb+1}=\alpha^{k,\kb}_{m,n+1,\nb+1}\,,\qquad \forall m,n,\nb\ge 0\,, \qquad k,\kb\ge 1\,.
	\end{equation}
	The solution to these constraints is simply
	\begin{equation}
	\label{all alpha}
	\alpha^{k,\kb}_{m,n,\nb}=\alpha^{k,\kb}_{1,0,0}\,, \qquad (m,n,\nb)\neq(0,0,0)\,.
	\end{equation}
	Thus at fixed $k,\kb$ all coefficients equal $\alpha^{k,\kb}_{1,0,0}$, except for the leading order coefficient $\alpha^{k,\kb}_{0,0,0}$ which is left unconstrained at this point. 
	
	Invariance under $L_1,\Lb_1$ then yields the recursion relations
	\begin{equation}
	\label{eq:recursion}
	\begin{split}
	(2h'+k+m+n-1)\alpha^{k,\kb}_{m,n,\nb}-(1+k) \alpha^{k+
		1,\kb}_{m,n,\nb}&=(2h_3+2k+m+n-2)\alpha^{k,\kb}_{m,n+1,\nb}\,,\\
	(2\hb'+\kb+m+\nb-1)\alpha^{k,\kb}_{m,n,\nb}-(1+\kb) \alpha^{k,\kb+
		1}_{m,n,\nb}&=(2\hb_3+2\kb +m+\nb-2)\alpha^{k,\kb}_{m,n,\nb+1}\,,
	\end{split}
	\end{equation}
	where we used the commutation relations
	\begin{equation}
	[L_1,(P_{-1,-1})^m]=m (P_{-1,-1})^{m-1}P_{0,-1}\,,  \quad [L_1,(L_{-1})^n]=2(L_{-1})^{n-1}\left(nL_0+\binom{n}{2}\right)\,.
	\end{equation}
	For $(m,n,\nb)\neq (0,0,0)$ and using \eqref{all alpha}, equation \eqref{eq:recursion} yields the recursive relations
	\begin{equation}
	\label{eq:recursion 2}
	\begin{split}
	\left(2h'-2h_3-k+1\right)\alpha^{k,\kb}_{1,0,0}&=(1+k)\, \alpha^{k+1,\kb}_{1,0,0}\,,\\
	\left(2\hb'-2\hb_3-\kb+1\right)\alpha^{k,\kb}_{1,0,0}&=(1+\kb)\,\alpha^{k,\kb+1}_{1,0,0}\,,
	\end{split}
	\end{equation}
	which allow to solve $\alpha^{k,\kb}_{1,0,0}$ in terms of $\alpha^{1,1}_{1,0,0}$, 
	\begin{equation}
	\alpha^{k,\kb}_{1,0,0}=\frac{\alpha^{1,1}_{1,0,0} (-1)^{k+\kb}\, \Gamma(2h_3-2h'+k-1)\Gamma(2\hb_3-2\hb'+\kb-1)}{k! \kb!\Gamma(2\hb_3-2\hb')\Gamma(2h_3-2h') }\,.
	\end{equation}
	For $(m,n,\nb)=(0,0,0)$ and using again \eqref{all alpha}, equation \eqref{eq:recursion} instead yields 
	\begin{equation}
	\label{eq:recursion 3}
	\begin{split}
	(2h'+k-1)\alpha^{k,\kb}_{0,0,0}-(1+k) \alpha^{k
		+1,\kb}_{0,0,0}&=2(h_3+k-1)\alpha^{k,\kb}_{1,0,0}\,,\\
	(2\hb'+\kb-1)\alpha^{k,\kb}_{0,0,0}-(1+\kb) \alpha^{k
		,\kb+1}_{0,0,0}&=2(\hb_3+\kb-1)\alpha^{k,\kb}_{1,0,0}\,.
	\end{split}
	\end{equation}
	These are recurrence relations for $\alpha^{k,\kb}_{0,0,0}$ that can be written in closed form. To do so, we first define $\tilde{\alpha}^{k,\kb}_{0,0,0}\equiv \alpha^{k,\kb}_{0,0,0}-\alpha^{k,\kb}_{1,0,0}$. Subtracting \eqref{eq:recursion 2} and \eqref{eq:recursion 3} then yields the recursion relation,
	\begin{equation}
	\begin{split}
	(1+k)\, \tilde \alpha^{k+1,\kb}_{0,0,0}&=(2h'+k-1)\, \tilde \alpha^{k,\kb}_{0,0,0}\,,\\
	(1+\kb)\, \tilde \alpha^{k,\kb+1}_{0,0,0}&=(2\hb'+\kb-1)\, \tilde \alpha^{k,\kb}_{0,0,0}\,,
	\end{split}
	\end{equation}
	whose solution is
	\begin{equation}
	\tilde{\alpha}^{k,\kb}_{0,0,0}=\tilde{\alpha}^{1,1}_{0,0,0}\, \frac{\Gamma(2h'+k-1)\Gamma(2\hb'+\kb-1)}{k!\kb!\Gamma(2h')\Gamma(2\hb')}\,.
	\end{equation}
	Eventually the free data is given by $\alpha^{1,1}_{0,0,0}$ and $\alpha^{1,1}_{1,0,0}$. The latter is what we called $\beta_1$ in \eqref{OPE ansatz}, while former simply corresponds to a normalization of the operator $O_3$ and can therefore be set to $\alpha^{1,1}_{0,0,0}=1$ (or equivalently $\tilde \alpha^{1,1}_{0,0,0}=1-\beta_1$) without loss of generality. It can be checked that at the lowest levels these equations agree with the solutions \eqref{OPE coefficients 1}-\eqref{OPE coefficients 2} (with $\gamma_3=0$). The simple form of the coefficients \eqref{all alpha} at fixed $k,\kb$ shows that the corresponding OPE is essentially a sum over Taylor expansions. We can therefore write a finite version of the OPE as
	\begin{equation}
	O_1(\x_1)O_2(\x_2)\sim z^{h}_{12}\zbar^{\hb}_{12}\sum^\infty_{k,\kb=1}\tilde{\alpha}^{k,\kb}_{0,0,0}(P_{-k,-\kb} O_{3'})(\x_2)+z^{h}_{12}\zbar^{\hb}_{12}\sum^\infty_{k,\kb=1}\alpha^{k,\kb}_{1,0,0}(P_{-k,-\kb} O_{3'})(\x_1),
	\end{equation}
	with the definition of descendant fields given in \eqref{descendant fields}.
	
	\subsubsection*{Chiral OPE}
	We note that the restriction $a=0$ comes as a result of not including $O_{3'}$ itself in the OPE but only its descendants. In order to have $a\neq 0$ and thus a time-dependent structure function $f_{123}$, it is necessary to include $O_{3'}$ as well as all other primary ancestors of $O_{3}$. While the structure function corresponding to $c_0\neq 0$ in \eqref{f12k} can be time-independent by restricting to the particular case $a=0$, this possibility is absent for the other OPE branches. For instance for the `chiral' OPE branch corresponding to $c_2\neq 0$ in \eqref{f12k}, we need to consider $O_{3'}$ and all other primary ancestors in order to satisfy the constraints imposed by Poincar\'e symmetry, with an OPE of the form 
	\begin{equation}
	\begin{split}
	O_1(\x)O_2(0)\sim \delta(\zbar)\, z^{h-\bar{h}-1} &\Big( [ancestors]+u^{2\bar h+1}[\beta' O_{3'}+z\,  \alpha'  L_{-1} O_{3'}+z^2...\, ]\\
	&+ u^{2\bar h+2}[\beta\, O_3+z\,  \alpha  L_{-1}O_3+z^2...\,]+u^{2\bar h+3}\,...\Big)\,,
	\end{split}  
	\end{equation}
	where $h, \bar{h}$ are given in \eqref{h definition}.
	Invariance under $P_{-1,0}\,,P_{0,0}\,,\bar L_1$ is guaranteed due to the presence of the delta distribution $\delta(\zbar)$. On the other hand, consistency with the action of $P_{0,-1}$ imposes
	\begin{equation}
	\alpha'=(2\bar h+2) \beta\,,
	\end{equation}
	while consistency with $ L_{-1}$ simply yields 
	\begin{equation}
	( h-\bar h-1)\beta=2 h_3  \alpha\,, \qquad ( h-\bar h-1)\beta'=(2 h_3-1)  \alpha'\,.    
	\end{equation}
	Thus we see that the tower of operators featuring $O_{3'}$ and its descendants that appear at order $u^{2\hb+1}$ have to be included, except in the fine-tuned case where $2\hb+2=0$ such that the structure function appearing in front of the primary operator $O_3$ is indeed time-independent. Similarly, the presence of a time-dependent structure function in front of the primary operator $O_{3'}$ requires to include its own parent in the OPE, and so on and so forth, such that all ancestors of $O_3$ are eventually included.  
	
	The expansion coefficients of this OPE can also be determined in closed form. Consider first the case where $2\bar h+2=0$ so that there is indeed a time-independent leading term, with the ansatz
	\begin{equation}
	\label{eq:deltaOPEansatz}
	O_1(\x)O_2(0)\sim \delta(\bar z) z^{h}\sum^\infty_{m,n=0}\alpha_{m,n}\frac{u^{m}}{m!}\frac{z^{n}}{n!} (P_{-1,-1})^m (L_{-1})^nO_3(0)\,,
	\end{equation}
	that is compatible with scale-covariance. 
	Due to the presence of the delta function, the only constraints on this ansatz come from invariance under $P_{0,-1}$ and $L_1$. They lead to the two recursion relations
	\begin{equation}
	(h+2h_1+n+m)\alpha_{m,n}=(2h_3+m+n)\alpha_{m,n+1}\,,\qquad \alpha_{m,n+1}=\alpha_{m+1,n}\,, 
	\end{equation}
	which can be solved as
	\begin{equation}
	\label{eq:OPEcoefficientsdelta}
	\alpha_{m,n}
	%=\alpha_{0,0}\frac{\Gamma[h_3+h_2-h_1]\Gamma(h_3-h_2+h_1+m+n)}{\Gamma(2h_3+m+n)}
	=\alpha\, B(h_3+h_2-h_1,h_3-h_2+h_1+m+n)\,,
	\end{equation}
	with $\alpha$ some overall normalisation.
	Note that we did not need to consider any BMS descendants in the ansatz \eqref{eq:deltaOPEansatz}. The reason for this is that the generators $L_1$ and $P_{0,-1}$, that allow one to move up and down the tower of descendants, commute with one another, similar to the case of $L_1$ and $\bar L_1$ in the case of a standard CFT. This is to be contrasted with \eqref{OPE ansatz}. There, the additional generators $P_{-1,0}\,, P_{0,0}\,, \bar{L}_1$ impose additional restrictions on the coefficients that do no allow non-trivial solutions without BMS descendants. In Section \ref{section 5.3} we will discuss a resummation of \eqref{eq:deltaOPEansatz} valid at finite $z$.
	
	Consider now the case where $k=2\bar h+2 \in \mathbb{N}$ is a positive integer. We can then easily adapt the above discussion by setting 
	\begin{equation}
	O_3=\partial^{k}_u O_4\qquad (\Delta_4=\Delta_3-k). 
	\end{equation}
	The OPE expansion \eqref{eq:deltaOPEansatz} can then be used with $O_4$ in place of $O_3$. Concretely we have
	\begin{equation}
	\label{eq:chiralantiderivative}
	\begin{split}
	&O_1(\x)O_2(0)\sim \delta(\bar z) z^{h}\sum^\infty_{m,n=0}\alpha'_{m,n}\frac{u^{m}}{m!}\frac{z^{n}}{n!} (P_{-1,-1})^m (L_{-1})^n\, O_4(0)\\
	&=\delta(\bar z) z^{h}\sum^{\infty}_{n=0}\frac{z^{n}}{n!}\left(\sum^{k-1}_{m=0}\alpha'_{m,n}\frac{u^{m}}{m!}(\partial^{-1}_u)^{k-m}+\sum^\infty_{m=k}\alpha'_{m,n}\frac{u^{m}}{m!}(P_{-1,-1})^{m-k}\right)(L_{-1})^nO_3(0)\,,
	\end{split}
	\end{equation}
	where $(\partial^{-1}_u)$ is an anti-derivative operator. The coefficients $\alpha'_{m,n}$ can be obtained from \eqref{eq:OPEcoefficientsdelta} upon replacing $h_3\rightarrow h_3-k$.   
	
	For all other values of $2\bar h+2$, we need to include an infinite number of parents to complete the OPE as was pointed out above. To tackle this task, it might be easier to use the OPE block construction of section~\ref{section 5.3}. We give further comments there.
	
	\subsubsection*{Ultralocal OPE}
	Finally, we consider the ultra-local OPE branch corresponding to $c_1 \neq 0$ in \eqref{f12k}. In this case, the presence of both delta functions $\delta(z)\delta(\zbar)$ is such that it is not necessary to explicitly include parent operators, and we can work with the ansatz
	\begin{equation}
	\label{doubledeltaOPE ansatz}
	O_1(\x)O_2(0)\sim \frac{\delta(z)\delta(\zbar)}{u^{\Delta_1+\Delta_2-\Delta_3-2} }\Big[O_3+u ( \beta_1 P_{-1,-1}O_3+\beta_2 L_{-1}\bar L_{-1} O_{3'})+\ldots\Big]\,.
	\end{equation}
	Note that we necessarily have $J_1+J_2=J_3$.
	From invariance under  $P_{-1,0}\,,P_{0,-1}\,,P_{0,0}$ we simply find
	\begin{equation}
	\beta_2=0\,,
	\end{equation}
	and $\beta_1$ is again arbitrary. Note that this is consistent since $P_{-1,-1}O_3$ could be itself considered as a distinct primary operator with the same OPE as $O_3$ but shifted weights.
	
	\subsubsection*{Casimir constraint and two-particle representations} 
	Let us now comment on the necessity of the `massive' terms for consistency of the proposed OPE expansion. Indeed without these we would be  essentially proposing that a tensor product of two massless representations can be decomposed into massless representations, while it is well-known that massive representations also appear in general \cite{Barut:1986dd}. Put very simply, the total momentum of a pair of massless particles is not null,
	\begin{equation}
	\label{total momentum}
	(p_1+p_2)^2=2\, p_1 \cdot p_2 \propto \omega_1 \omega_2 |x_{12}|^2\,,    
	\end{equation}
	unless the particles are \textit{exactly} colinear or at least one of the momenta is zero (in which case we should speak of zero-momentum rather than massless representation).  
	In terms of carrollian operators, using the general identity
	\begin{equation}
	\left[AB,O_1O_2\right]=[A,[B,O_1]]O_2+O_1[A,[B,O_2]]+[A,O_1][B,O_2]+[B,O_1][A,O_2]\,,
	\end{equation} 
	we can evaluate the action of the quadratic Casimir operator $\mathcal{C}_2=-(HK+KH)+2B^i B_i$
	on the left-hand side of \eqref{Carrollian OPE}, yielding
	\begin{equation}
	\label{Casimir product}
	\begin{split}
	\left[\mathcal{C}_2, O_1 O_2\right]&=-2[H,O_1][K,O_2]-2[K,O_1][H,O_2]+4[B^i,O_1][B_i,O_2]\\
	&=2(x_1^2-2x_1 \cdot x_2+x_2^2)\partial_u O_1\partial_u O_2=2|x_{12}|^2\partial_u O_1\partial_u O_2\,.
	\end{split}
	\end{equation}
	This gives the total invariant mass of the product $O_1O_2$, which is indeed the same as \eqref{total momentum} modulo a Fourier transform. While \eqref{Casimir product} is generically nonzero, acting with  $\mathcal{C}_2$ on the right-hand side of \eqref{Carrollian OPE} would yield a strict zero if no massive operators were included, since $[\mathcal{C}_2,O]=0$ for any single-particle carrollian conformal field $O$. Thus massive operators need to be included in the carrollian OPE \eqref{carrollian OPE block}. In section~\ref{subsection two particle} we have constructed a local field $\psi(\x)$ with nonzero mass and part of an indecomposable multiplet $(\phi,\psi)$. Let us see how it can help resolve the situation, focusing on the contribution from a single one-particle field $O_k$ without loss of generality. For each such single-particle operator, we will need to add two multiplets $(\phi,\psi)$ and $(\phi',\psi')$, resulting in the OPE
	\begin{equation}
	\label{OPE block two particle}
	O_1(\x_1) O_2(\x_2) \approx f_{12k}(\x_{12})\, O_k(\x_2)+f_{12\psi}(\x_{12})\, \psi(\x_2)+f_{12\psi'}(\x_{12})\, \psi'(\x_2) +subl\,.
	\end{equation}
	Acting with $\mathcal{C}_2$ on the left-hand side results in \eqref{Casimir product}, and subsequently inserting \eqref{OPE block two particle} yields
	\begin{equation}
	\begin{split}
	&\left[\mathcal{C}_2, O_1(\x_1) O_2(\x_2)\right]\\
	&\approx 2|x_{12}|^2 \partial_{u_1} \partial_{u_2}\left(f_{12k}(\x_{12}) O_k(\x_2)+f_{12\psi}(\x_{12})\, \psi(\x_2)+f_{12\psi'}(\x_{12})\, \psi'(\x_2)\right)+subl\,.
	\end{split}
	\end{equation}
	On the other hand, acting with $\mathcal{C}_2$ on the right-hand side of \eqref{OPE block two particle}, and making use of \eqref{C2 psi} with $\beta=0$ for concreteness, yields 
	\begin{align}
	\left[\mathcal{C}_2, O_1(\x_1) O_2(\x_2)\right]&\approx 2 f_{12\psi}(\x_{12})\, \partial_u^2\phi(\x_2)+2 f_{12\psi'}(\x_{12})\, \partial_u^2\phi'(\x_2)+subl.\,,
	\end{align}
	Consistency at leading order in $x_{12}^i \sim 0$ can thus be established by setting
	\begin{equation}
	\partial_u \phi'(\x)=O_k(\x)\,, \qquad f_{12\psi'}(\x)=|x|^2 \partial_u f_{12k}(\x)\,.
	\end{equation}
	and
	\begin{equation}
	\partial_u^2 \phi(\x)=O_k(\x)\,, \qquad f_{12\psi}(\x)=-|x|^2 \partial_u^2 f_{12k}(\x)\,,
	\end{equation}
	This means in particular that $\phi$ and $\phi'$ must be identified with the first two parent primaries of $O_k$. Since $\Delta_\psi=\Delta_\phi+2$, we also infer the scaling dimensions 
	\begin{equation}
	\Delta_\psi=\Delta_k\,, \qquad \Delta_{\psi'}=\Delta_k+1\,.    
	\end{equation}
	The structure functions $f_{12\psi}$ and $f_{12\psi'}$ can be seen to have the corresponding scaling weights. In summary, we see that the indecomposable multiplets $(\phi,\psi)$ are precisely of the type needed to satisfy the quadratic Casimir constraint, in relation to two-particle exchange in the context of massless particle scattering. We leave the detailed study of their OPE blocks to future work.
	
	\subsection{Holomorphic coincidence limit and colinear factorisation}
	\label{section 5.2}
	We now study the holomorphic coincidence limit, motivated by its relation with colinear factorisation of massless scattering amplitudes given in \cite{Mason:2023mti}. Since in this case we allow for finite separations $u_{12}\neq 0$ and $\zbar_{12}\neq 0$, the question naturally arises as to where should we place the operators $O_3$ appearing in the resulting OPE. Arguably the most sensible ansantz involves integrating its position over the intervals separating the two insertions, namely
	\begin{equation}
	\label{integral OPE ansatz}
	O_1(\x_1) O_2(\x_2) \stackrel{z_{12}\sim 0}{\approx} \int_0^1 dt \int_0^1 ds\, F(\x_{12};t,s)\, O_3(u_2+t u_{12},z_2,\zbar_2+s \zbar_{12})\,,
	\end{equation}
	which can be readily checked to be consistent with carrollian translations generated by $H, L_{-1}, \bar L_{-1}$. Setting $\x_2=0$ without loss of generality, we have
	\begin{equation}
	O_1(\x)O_2(0)\stackrel{z\sim 0}{\approx} z^\delta \int^1_0\, dt \int_0^1 ds\, F(u,\zbar;s,t)O_3(ut,0,s\bar{z})\,,
	\end{equation}
	where we have also assumed a leading power-law behavior $z^\delta$ with an exponent $\delta$ to be determined. Note that we do not need to explicitly write derivatives $\partial_u$ and $\bar \partial$ acting on $O_3$ since one can use integration by parts and redefine the function $F(u,\zbar;s,t)$ to reabsorb them. 
	
	We will impose now the constraints implied by conformal carrollian symmetry. Acting with $L_0, \bar{L}_0$ yields the conditions
	\begin{equation}
	\label{L0 constraint}
	\begin{split}
	u\partial_u F+2(h_1+h_2-h_3+\delta)F&=0\,,\\
	u\partial_u F+2\zb \bar{\partial}F+2(\hb_1+\hb_2-\hb_3)F&=0\,,
	\end{split}
	\end{equation}
	while acting with $P_{-1,0}$ requires
	\begin{equation}
	\int^1_0\, dt \int_0^1 ds\left(\partial_u F\, O_3+u^{-1}(t-s)F\, \partial_t O_3 \right)=0\,.
	\end{equation}
	An immediate solution to the latter is given by 
	\begin{equation}
	\label{eq:collinsol}
	F=\delta(t-s) f(\bar{z};t)\,,
	\end{equation}
	such that the first equation in \eqref{L0 constraint} fixes
	\begin{equation}
	\delta=h_3-h_2-h_1\equiv h\,.
	\end{equation}
	Consistency with the action of $\bar{L}_1$ then implies
	\begin{equation}
	\int^1_0\, dt \left(\bar{z} \bar{\partial}f\, O_3 +t(1-t) f \frac{d}{d t}O_3+2 O_3\, f(\bar{h}_1-t \bar{h}_3)\right)=0\,,
	\end{equation} 
	which, after integration by parts, yields the differential equation
	\begin{equation}
	\frac{d}{dt}\left(f\, t(1-t)\right)-2f(\bar{h}_1-t \bar{h}_3)+(\bar{h}_1+\bar{h}_2-\bar{h}_3)f=0\,.
	\end{equation}
	Together with the constraints \eqref{L0 constraint}, the solution is given by
	\begin{equation}
	f(\bar{z};t)=c_{123}\,\zb^{\hb_3-\hb_2-\hb_1}\,t^{\hb_3-\hb_2+\hb_1-1}(1-t)^{\hb_3+\hb_2-\hb_1-1}\,,
	\end{equation}
	Note that the boundary contributions arising from integrating by parts vanish only if $\bar h=\hb_3-\hb_2-\hb_1>0$, which we therefore have to assume. In summary, we found the leading OPE term
	\begin{equation}
	\label{eq:holomorphicOPE}
	O_1(\x)O_2(0)\stackrel{z\sim 0}{\approx} c_{123}\,z^h\,\zb^{\hb}\int^1_0 dt \,t^{\hb_3-\hb_2+\hb_1-1}(1-t)^{\hb_3+\hb_2-\hb_1-1} O_3(tu,0,t\zbar)\,.
	\end{equation}
	Consistency with the action of $P_{0,-1}\,,P_{0,0}\,,L_1$ should determine the subleading terms in $z\sim 0$ involving descendant operators. 
	
	We are now in a position to discuss the carrollian OPE obtained from collinear factorisation of massless tree-level amplitudes presented in \cite{Mason:2023mti}, which is in fact contained in \eqref{eq:holomorphicOPE}. In that case, we expect a leading $z^{-1}$ pole from the collinear limit which fixes $h_3=h_2+h_1-1$. To compare with the formula in \cite{Mason:2023mti}, we further set $\Delta_{1,2}=1\,, \Delta_3=1+p$, such that
	\begin{equation}
	\label{Mason et al}
	O_{1,J_1}(\x)O_{1,J_2}(0)\stackrel{z\sim 0}{\approx} c_{123}\,z^{-1}\,\zb^{p}\int^1_0 dt \,t^{J_2-J_3-1}(1-t)^{J_1-J_3-1} O_{1+p,J_3}(tu,0,t\zbar),
	\end{equation}
	with $p=J_{1}+J_2-J_3-1>0$ due to the above requirement of vanishing boundary terms. The resulting expression \eqref{Mason et al} is identical to the one obtained in \cite{Mason:2023mti}, provided we express the primary operator $O_{1+p,J_3}$ in terms of its $p$-th `ancestor' $O_{1,J_3}$ via $O_{1+p,J_3}=(\partial_u)^p O_{1,J_3}$.
	
	One could further expand \eqref{Mason et al} in powers of $u$ and $\zbar$, and explicitly perform the $t$-integral. The resulting power series can be found in \cite{Mason:2023mti,Ruzziconi:2024kzo}. In what follows, we display the first few terms and discuss an apparent inconsistency with the general OPE \eqref{OPE ansatz} of the previous section. 
	
	\subsubsection*{Connection with the coincidence limit} 
	Let us expand the holomorphic OPE \eqref{eq:holomorphicOPE} to first order in $\zbar$ and $u$, and check that it is indeed of the general form found in the previous subsection. We find
	\begin{equation}
	\label{expanded holomorphic OPE}
	\begin{split}
	O_1(\x) O_2(0)&\approx c_{123}\, B(\bar h_3-\bar h_2+\bar h_1,\bar h_3+\bar h_2-\bar h_1) z^h \zbar^{\bar h} \\
	&\times \left(O_3(0)+\frac{\bar h_3-\bar h_2+\bar h_1}{2\bar h_3} (\zbar \partial_{\zbar}+u \partial_u ) O_3(0) \right)+...\,,
	\end{split}
	\end{equation}
	with the Euler beta function given by
	\begin{equation}
	B(a,b)\equiv \int_0^1 dt\, t^{a-1} (1-t)^{b-1}=\frac{\Gamma[a]\Gamma[b]}{\Gamma[a+b]}\,.
	\end{equation}
	Comparing with \eqref{OPE ansatz}, we identify the OPE coefficients
	\begin{equation}
	\bar \alpha_1=\beta_1=\frac{\bar h_3-\bar h_2+\bar h_1}{2\bar h_3}\,, \qquad \bar \alpha_2=\beta_2=0\,,
	\end{equation}
	consistently with \eqref{OPE coefficients 1} and the second equation in \eqref{OPE coefficients 2}. From \eqref{OPE coefficients 2} we can also directly determine the coefficients $\gamma_2, \gamma_4, \bar \gamma_4$. 
	
	However, there is an inconsistency between the holomorphic OPE \eqref{eq:holomorphicOPE} and the OPE \eqref{OPE ansatz} when going to subsubleading orders. Indeed, assuming the validity of the latter, the parameter $\gamma_4=(\bar h'-\bar h_3)\beta_1$ is generically nonzero and implies the appearance of BMS descendants of both $O_3$ and its parent $O_{3'}$ at subsubleading orders. Looking at \eqref{full OPE} and \eqref{all alpha}, we indeed see that a nonzero $\gamma_4=\alpha^{1,2}_{1,0,0}$ implies in particular the appearance of $P_{-2,-1}O_3$ at order $u\zbar$ and $L_{-1}P_{-2,-1}O_{3'}$ at order $\zbar^2$. Obviously such terms are not produced when expanding \eqref{eq:holomorphicOPE} to these orders, which would signal its failure to satisfy all Poincar\'e constraint. It is likely that adding descendants of higher parents to the ansatz \eqref{OPE ansatz} would resolve this apparent tension, but we leave this study to future endeavors. Moreover, it can also happen that some terms in the OPE that feature BMS descendants actually drop out when evaluated inside correlation functions, as a result of \eqref{zn delta} for instance. 
	
	Relatedly, we should also keep in mind that the colinear factorisation of massless scattering amplitudes used in \cite{Mason:2023mti} to derive \eqref{Mason et al} only holds to first order in the colinear expansion $p_1\cdot p_2 \sim 0$. This pairs well with the fact that the first orders \eqref{expanded holomorphic OPE} agree with the carrollian OPE \eqref{OPE ansatz}. Investigation of the subleading orders in the colinear expansion, discussed in \cite{Nandan:2016ohb,Banerjee:2020zlg,Adamo:2022wjo,Ren:2023trv}, and their agreement with the subleading terms in the carrollian OPE constitutes an interesting open problem.

	\subsection{OPE blocks}
	\label{section 5.3}
	We now turn to the discussion of OPE blocks, first introduced in the context of standard conformal field theory in \cite{Czech:2016xec}. Their purpose is to resum the OPE \eqref{standard OPE coincident} such as to produce a formula valid for finite separation $\x_{12}\neq 0$. We adapt the discussion to the carrollian setup, assuming an ansatz of the form 
	\begin{equation}
	\label{carrollian OPE block}
	O_1(\x_1) O_2(\x_2)\sim \int_{\mathcal{D}(\x_1,\x_2)} d^3\x\, F_{12k}(\x_1,\x_2,\x)\, O_k(\x)\,,
	\end{equation}
	where $\mathcal{D}(\x_1,\x_2)$ is some domain of integration which depends on the operator insertions, and $F_{12k}(\x_1,\x_2,\x)$ is some three-point function, both to be determined.
	Under coordinate transformations \eqref{complex coordinate transformations}, the integration measure transforms like
	\begin{equation}
	d^3\x'=\left(\frac{\partial z'}{\partial z} \right)^{3/2} \left(\frac{\partial \zbar'}{\partial \zbar} \right)^{3/2} d^3\x\,,
	\end{equation}
	such that, using the transformation law \eqref{conformal transfo} for the operator $O_k$,
	\begin{equation}
	\label{O1 O2 prime}
	\begin{split}
	O'_1(\x_1') O'_2(\x_2')&\sim \int_{\mathcal{D}(\x_1',\x_2')} d^3\x'\, F'_{12k}(\x_1',\x_2',\x')\, O'_k(\x')\\
	&=\int_{\mathcal{D}'(\x_1',\x_2')} d^3\x\, \left(\frac{\partial z'}{\partial z} \right)^{3/2-h_k} \left(\frac{\partial \zbar'}{\partial \zbar} \right)^{3/2-\bar h_k} F'_{12k}(\x_1',\x_2',\x')\, O_k(\x)\,.
	\end{split}
	\end{equation}
	On the other hand using the transformation of the operators $O_1(\x_1) O_2(\x_2)$, we must also have
	\begin{equation}
	\label{O1 O2 prime}
	\begin{split}
	O'_1(\x_1') O'_2(\x_2')&\sim \left(\frac{\partial z_1'}{\partial z_1} \right)^{-h_1}\left(\frac{\partial \zbar_1'}{\partial \zbar_1} \right)^{-\bar h_1} \left(\frac{\partial z_2'}{\partial z_2} \right)^{-h_2} \left(\frac{\partial \zbar_2'}{\partial \zbar_2} \right)^{-\bar h_2}\\
	&\times \int_{\mathcal{D}(\x_1,\x_2)} d^3\x\, F_{12k}(\x_1,\x_2,\x)\, O_k(\x)\,.
	\end{split}
	\end{equation}
	For consistency $F_{12k}$ must therefore behave like a carrollian three-point function
	\begin{equation}
	\label{eq:shadow3pt}
	F_{12k}(\x_1,\x_2,\x)=\langle O_1(\x_1) O_2(\x_2) \tilde O_k(\x) \rangle\,,
	\end{equation}
	where the fictitious \textit{shadow operator} $\tilde O_k$ has dimension $\tilde h_k=3/2-h_k$ and $\tilde{\bar h}_k=3/2-\bar h_k$, or equivalently $\tilde \Delta=3-\Delta$ and $\tilde J=-J$. In addition, the domain of integration must be invariant under carrollian conformal transformations,
	\begin{equation}
	\mathcal{D}'(\x_1',\x_2')=\mathcal{D}(\x_1,\x_2)\,.
	\end{equation}
	For the spatial domain of integration, we can take the same one as in CFT$_2$, since carrollian conformal transformations \eqref{complex coordinate transformations} act just as $2d$ conformal transformations on the celestial sphere. This is a diamond in the $(z,\zbar)$-plane, with edges given by $(z_1,\zbar_1)$ and $(z_2,\zbar_2)$ \cite{Czech:2016xec}. For the time domain, we could integrate $u$ over the whole real axis for instance, or define it as a closed contour in complex $u$-plane. When we turn to carrollian amplitudes and the specific examples discussed in Section \ref{section 6}, we will also see that the Heaviside functions coming from energy positivity (see equations \eqref{eq:3ptwithTheta} and \eqref{4-point general}) determine a particular choice of integration range along $u$. In the following, we will leave it unspecified until needed.
	
	It is instructive to see how carrollian OPE blocks might be related to the celestial OPE block constructed in \cite{Guevara:2021tvr}. The latter are given by
	\begin{align}
	O_{1+i\nu_1}(\vec x_1)\, O_{1+i\nu_2}(\vec x_2)\sim \int_{-\infty}^\infty d\nu \int d^2\vec x\, \langle O_{1+i\nu_1}(\vec x_1) O_{1+i\nu_2}(\vec x_2) \tilde O_{1-i\nu}(\vec x) \rangle\, O_{1+i\nu}(\vec x)\,, 
	\end{align}
	where all operator are SL(2,$\mathbb{C}$) primary fields of the principal continuous series. Indeed such operators provide a basis for decomposing both massless and massive one-particle states \cite{Iacobacci:2024laa}. In order to obtain a statement for carrollian operators, we apply the transformation \cite{Donnay:2022wvx}
	\begin{equation}
	O_\Delta(\x)=\int_{-\infty}^{\infty} d\nu\, \frac{\Gamma[\Delta-1-i\nu]}{(u\mp i0^+)^{\Delta-1-i\nu}}\,  O_{1+i\nu}(\vec x)\,,
	\end{equation}
	such that we obtain
	\begin{align}
	O_{\Delta_1}(\x_1) O_{\Delta_2}(\x_2)=\int d^2\vec x\, \int_{-\infty}^\infty d\nu d\nu' \delta(\nu-\nu')\, \langle O_{\Delta_1}(\x_1) O_{\Delta_2}(\x_2) \tilde O_{1-i\nu}(\vec x) \rangle\, O_{1+i\nu'}(\vec x)\,.    
	\end{align}
	We can use the following representation of the delta distribution,
	\begin{equation}
	4\pi \delta(\nu-\nu')=\int_{-\infty}^\infty du\, \frac{\Gamma[2-\Delta+i\nu]}{(u+i0^+)^{2-\Delta+i\nu}} \frac{\Gamma[\Delta-1-i\nu']}{(u-i0^+)^{\Delta-1-i\nu'}}\,, \qquad \Delta \in \mathbb{R}\,,
	\end{equation}
	such as to complete the change of basis from celestial to carrollian fields,
	\begin{align}
	O_{\Delta_1}(\x_1) O_{\Delta_2}(\x_2)=\int d^3\x\,  \langle O_{\Delta_1}(\x_1) O_{\Delta_2}(\x_2) \tilde O_{3-\Delta}(\x) \rangle\, O_{\Delta}(\x)\,.    
	\end{align}
	In this way we have formally recovered the carrollian OPE block discussed above.
	
	An important distinction compared to standard conformal field theory, is that there exists a variety of three-point functions for any given set of fields, as we discussed at length in section~\ref{subsection 3.2}. Each possible three-point function potentially defines an OPE block. Similarly, we have shown in section~\ref{section 5.1} that there exist different branches of OPEs in the coincident limits, and we expect that there is a correspondence with the various OPE blocks one can define. Let us show this explicitly.
	
	\subsubsection*{Ultralocal OPE} 
	We first aim to recover the ultralocal OPE~\eqref{doubledeltaOPE ansatz}. By inspection, it is clear that the relevant three-point function from section~\ref{subsection 3.2} should be \eqref{eq:ultralocal3pt}. Plugging it in \eqref{eq:shadow3pt} we thus have
	\begin{equation}
	\begin{split}
	O_1(\x_1)O_2(\x_2)&=c_{123}\int d^3\x_3 \frac{\delta^{(2)}(\vec x_{12})\delta^{(2)}(\vec x_{23})}{u^a_{12}u^b_{23}u^c_{31}}\, O_3(\x_3)\\
	&=c_{123}\, \delta^{(2)}(\vec x_{12})\int d u_3 \frac{1}{u^a_{12}u^b_{23}u^c_{31}}\, O_3(\x_3)\,,
	\end{split}
	\end{equation}
	where $a+b+c+1=\Delta_1+\Delta_2-\Delta_3$. The integration range for $\vec x_3$ is arbitrary as long as it includes the support of the delta function.
	Using the change of variables $u_3=u_2+t u_{12}$, we can further write
	\begin{equation}
	O_1(\x_1)O_2(\x_2)=\frac{c_{123}\delta(z_{12})\delta(\zbar_{12})}{u^{\Delta_1+\Delta_2-\Delta_3-2}_{12}}\int dt \frac{O_3(u_2+t u_{12},z_2,\zb_2)}{(-t)^b(-1+t)^c}\,.
	\end{equation}
	Expanding this in powers of $u_{12}$ reproduces \eqref{doubledeltaOPE ansatz}, with coefficients determined by the choice of integration range for $u_3$. %In particular, the integration limits $const,\pm \infty$ are all consistent with symmetries. 
	This pairs well with the fact that the normalization of $O_3$ and the coefficient $\beta_1$ in \eqref{doubledeltaOPE ansatz} are also arbitrary.
	
	\subsubsection*{Chiral OPE} 
	Aiming to recover the chiral OPE \eqref{eq:deltaOPEansatz}, the relevant three-point function would appear to be the chiral three-point function \eqref{generic 3-point function}. We thus write
	\begin{equation}
	\label{eq:chiralOPEblock}
	\begin{split}
	O_1(\x_1) O_2(\x_2)&=\int d^3\x_3 \frac{c_{123}\, \delta(\zbar_{12})\delta(\zbar_{23})}{ (z_{12})^a\, (z_{23})^b\, (z_{13})^c\, (F_{123})^d}\, O_3(\x_3)\\
	&=\frac{c_{123}\, \delta(\zbar_{12})}{z^{a+b+c+d-1}_{12}u^{d-1}_{12}}\int d t d s\frac{O_3(u_2+t u_{12},z_2+sz_{12},\zbar_2)}{(-s)^b(-1+s)^c(-s+t)^d}\,,
	\end{split}
	\end{equation}
	where we made the variable changes $u_3=u_2+tu_{12}$ and $z_3=z_2+s z_{12}$, and where $a,b,c,d$ are given in \eqref{d parameter} and \eqref{abc parameters} subject to the replacement $\Delta_3 \mapsto 3-\Delta_3$ and $J_3\mapsto -J_3$. It can be checked that the leading term in the expansion $u_{12}\,, z_{12}\sim 0$ agrees with that of \eqref{eq:deltaOPEansatz}.
	
	Consider now the special case $1-d=2\hb+2=0$ for which the leading $u$-dependence vanishes. We choose a contour in $u$, or equivalently $t$, that circles the pole in $(s-t)^{-1}$. Note that $F_{123}$ naturally contains an imaginary part that shifts the pole away from the real axis. Choosing furthermore the integration bounds $z_3\in(z_1,z_2)$ we have then, up to an (imaginary) prefactor that we reabsorb in $c_{123}$,
	\begin{equation}
	\begin{split}
	O_1(\x_1) O_2(\x_2)&= -\frac{c_{123}\, \delta(\zbar_{12})}{z^{a+b+c}_{12}} \int^{1}_0 ds\, \frac{O_3(u_2+s u_{12},z_2+sz_{12},\zbar_2)}{s^b(1-s)^c}\\
	&=-\frac{c_{123}\, \delta(\zbar_{12})}{z^{a+b+c}_{12}}\int^{1}_0ds \sum^\infty_{m,n=0} \frac{u^m_{12}}{m!}\frac{z^n_{12}}{n!} s^{m+n-b}(1-s)^{-c}\, \partial^m_{u_2}\partial^n_{z_2}O_3(\x_2)\\
	&=-\frac{c_{123}\, \delta(\zbar_{12})}{z^{a+b+c}_{12}}\sum^\infty_{m,n=0}\frac{u^m_{12}}{m!}\frac{z^n_{12}}{n!} B(m+n+1-b,1-c)\, \partial^m_{u_2}\partial^n_{z_2}O_3(\x_2).
	\end{split}
	\end{equation}
	Reabsorbing the leading term in the free coefficient $c_{123}$ and plugging in the value of the parameters $a,b,c$ (taking into account the above-mentioned shifts) we find
	\begin{equation}
	\label{eq:finalOPE3}
	O_1(\x_1)O_2(\x_2)=c'_{123}\, \delta(\zbar_{12}) z^{h}_{12}\sum^\infty_{m,n=0}\frac{u^m_{12}}{m!}\frac{z^n_{12}}{n!} \frac{\Gamma(2h_3)\Gamma(h+2h_1+m+n)}{\Gamma(h+2h_1)\Gamma(2h_3+m+n)}\, \partial^m_{u_2}\partial^n_{z_2}O_3(\x_2),
	\end{equation}
	in perfect agreement with the OPE given in \eqref{eq:deltaOPEansatz}-\eqref{eq:OPEcoefficientsdelta}. Based on the above it is natural to take  \eqref{eq:chiralOPEblock} with an appropriately chosen $u$-contour as defining the chiral OPE for arbitrary values of the weights.
	
	In writing \eqref{eq:chiralOPEblock} we could have chosen to include the Heaviside functions which define the 3-point amplitudes. Let us discuss briefly how their inclusion influences the resulting OPE. All operators carry now an additional label $\eta_i$ that distinguishes ingoing from outgoing operators. As we will see, the OPE can depend on this additional `flavor' label. As before, we write
	\begin{equation}
	\begin{split}
	&O^{\eta_1}_1(\x_1) O^{\eta_2}_2(\x_2)=c_{123}\int d^3\x_3 \langle O^{\eta_1}_1O^{\eta_2}_2\tilde{O}^{-\eta_3}_3\rangle O^{\eta_3}_3\\
	&=c_{123}\int d^3\x_3 \frac{\delta(\zbar_{12})\delta(\zbar_{23})}{ (z_{12})^a\, (z_{23})^b\, (z_{13})^c\, (F_{123})^d}\, \Theta\left(-\frac{z_{13}}{z_{23}}\eta_1\eta_2\right)\Theta\left(-\frac{z_{12}}{z_{23}}\eta_1\eta_3\right)O^{\eta_3}_3(\x_3)\,,
	\end{split}
	\end{equation}
	where we inserted the three-point amplitude \eqref{eq:3ptwithTheta} to define the block, and $a,b,c,d$ are again subject to the replacement $\Delta_3 \mapsto 3-\Delta_3$ and $J_3\mapsto -J_3$. Note that we take the shadow operator $\tilde{O}_3$ to have opposite in/out label compared to $O_3$.
	Going through the same steps as above but leaving the integration range of $z_3$ unspecified for the moment we get to
	\begin{equation}
	\begin{split}
	\label{eq:chiralwitheta}
	O^{\eta_1}_1(\x_1) O^{\eta_2}_2(\x_2)= -\frac{c_{123}\, \delta(\zbar_{12})}{z^{a+b+c}_{12}} \int ds\, &\sum^\infty_{m,n=0} \frac{u^m_{12}}{m!}\frac{z^n_{12}}{n!} s^{m+n-b}(1-s)^{-c}\, \partial^m_{u_2}\partial^n_{z_2}O^{\eta_3}_3(\x_2)\\
	&\qquad \times \Theta\left(\frac{1-s}{s}\eta_1\eta_2\right)\Theta\left(\frac{1}{s} \eta_1\eta_3\right).
	\end{split}
	\end{equation}
	For a given choice of in/out configuration the Heaviside functions determine the integration range. In particular, we have
	\begin{equation}
	\begin{aligned}
	&\eta_1=\eta_2=\eta_3 &\qquad s&\in(0,1)\,,\\
	&\eta_1=-\eta_2=-\eta_3 &\qquad s&\in(-\infty,0)\,,\\
	&\eta_1= -\eta_2=\eta_3 &\qquad s&\in(1,\infty)\,.
	\end{aligned}
	\end{equation}
	We see from the first equation that two ingoing operators can only fuse into another ingoing operator in which case we exactly recover the previous result \eqref{eq:finalOPE3}. On the other hand, the OPE of operators with opposite $\eta$-label can consist of two blocks for each possible $\eta$-label of $O_3$. Note that all resulting integrals lead to the same OPE expansion (up to an overall constant) that is consistent with \eqref{eq:deltaOPEansatz} and \eqref{eq:OPEcoefficientsdelta}. 
	
	%For the second case, we have the integral
	%\begin{align}
	%    \int^{\infty}_1ds s^{m+n-b}(1-s)^{-c}&=\frac{_{2}F_1(c,c+b-1-m-n,c+b-m-n;1)}{(-1)^{c}(c+b-1-m-n)}=(-1)^{-c}B(1-c,b+c-1-m-n)\\
	%    &=B(1-c,m+n+1-b)(-1)^{-c+1}\frac{\sin b\pi}{\sin(b+c)\pi}
	%\end{align}
	%and for the third case
	%\begin{align}
	%    \int^{0}_{-\infty}ds s^{m+n-b}(1-s)^{-c}&=(-1)^{m+n-b}\int^{\infty}_0 s^{m+n-b}(1+s)^{-c}=(-1)^{m+n-b}B(m+n+1-b,c+b-1-m-n).\\
	%    &=B(1-c,m+n+1-b)(-1)^{-b+1}\frac{\sin c\pi}{\sin(b+c)\pi}.
	%\end{align}
	
	\section{Realisation of OPEs in correlators and amplitudes}
	\label{section 6}
	
	In order to exemplify and check the relevance of the carrollian OPEs constructed in the previous section, we investigate their realisation within the carrollian correlation functions of Section~\ref{section 3} and carrollian MHV amplitudes of Section~\ref{section 4}. Note in the latter case, the presence of Heaviside distributions associated with the positivity of the particles' energies will unveil the realisation of a different OPE branch.  
	
	\subsection{3-point correlators and amplitudes} 
	
	\subsubsection*{Correlators}
	We start with the 3-point correlator given in \eqref{generic 3-point function}. In order to get a definite expression, the order in which the OPE limit $\x_{12} \to 0$ is taken must be specified. Since the 3-point correlator \eqref{generic 3-point function} contains a delta distribution $\delta(\zbar_{12})$, its argument $\zbar_{12}$ is necessarily the smallest parameter in the game. Thus let us choose the order of limit
	\begin{equation}
	\zbar_{12} \leq z_{12} \leq u_{12} \ll 1\,.
	\end{equation}
	In this case, we have
	\begin{equation}
	F_{123} \sim u_{12} z_{23}\,,
	\end{equation}
	such that
	\begin{equation}
	\langle O_1(\x_1) O_2(\x_2) O_3(\x_3) \rangle \sim c_{123} \frac{z_{12}^{\Delta_3-J_1-J_2-2} \delta(\zbar_{12})}{u_{12}^{2(\bar h_1+\bar h_2+\bar h_3-2)}} \delta(\zbar_{23}) z_{23}^{-2h_3}\,.
	\end{equation}
	If this limit is controlled by an OPE,  and calling $O_4$ the dominant exchanged primary whose quantum numbers must be determined, then we should be able to recast this formula in the form
	\begin{equation}
	\label{OPE ?}
	\langle O_1(\x_1) O_2(\x_2) O_3(\x_3) \rangle \stackrel{?}{\sim} f_{124}(\x_{12}) \langle O_4(\x_2) O_3(\x_3) \rangle\,,
	\end{equation}
	with $f_{124}(\x_{12})$ of the form \eqref{f12k}.
	This is indeed the case if the quantum numbers of $O_4$ are given by
	\begin{equation}
	h_4= h_3\,, \qquad \bar h_4=1-\bar h_3\,,  
	\end{equation}
	or equivalently
	\begin{equation}
	\Delta_4=J_3+1\,, \qquad J_4=\Delta_3-1\,,
	\end{equation}
	with $f_{124}(\x_{12})$ realising the $\delta(\zbar)$-branch of \eqref{f12k}, and $\langle O_4 O_3 \rangle$ given by the chiral two-point function \eqref{chiral 2-point function}.
	
	\subsubsection*{Amplitudes}
	We then turn to the OPE limit of the 3-point carrollian amplitude \eqref{eq:3ptwithTheta}, which we display again here for convenience,
	\begin{equation}
	\label{3-point amplitude OPE}
	\langle O_1 O_2 O_3 \rangle=\frac{\delta(\zbar_{12})\delta(\zbar_{23})}{ (z_{12})^a\, (z_{23})^b\, (z_{13})^c\, (F_{123})^d}\, \Theta\left(-\frac{z_{13}}{z_{23}}\eta_1\eta_2\right)\Theta\left(\frac{z_{12}}{z_{23}}\eta_1\eta_3\right),\nonumber
	\end{equation}
	withs $a,b,c,d$ given in equations \eqref{d parameter}-\eqref{abc parameters}.
	In the limit $z_{12}\rightarrow 0$, the support of the Heaviside distributions in the $z_{23}$-plane is vanishing  away from $z_{23}=0$. Hence it will be nontrivial as a distribution in the variable $z_{23}$ only if it becomes proportional to a delta distribution $\delta(z_{23})$. Let us see how this happens, by writing
	\begin{equation}
	\label{eq:3pttrick}
	\langle O_1 O_2 O_3 \rangle=\int \dd x\, \delta(x-z_{23})\, \langle O_1 O_2 O_3 \rangle= -z_{12} \int d s\, \delta(z_{23}+s z_{12})\langle O_1 O_2 O_3 \rangle\,,
	\end{equation}
	where we made the change of variables $z_{23}=-s z_{12}$ in the second step. As we will see momentarily, the $s$-integral converges so that these manipulations are meaningful in the sense of distributions in the limit $z_{12}\rightarrow 0$.
	We have then 
	\begin{align}
	\nonumber
	&\langle O_1 O_2 O_3 \rangle\\
	&=(-1)^{b+d-1}\frac{\delta(\zbar_{12})\delta(\zbar_{23})}{(z_{12})^{a+b+c+d-1}}  \int \frac{d s\, \delta(z_{23}+s z_{12})}{  s^{b} (1-s)^{c}(s u_{12}+u_{23})^{d}}\, \Theta\left(\frac{(1-y)\eta_1\eta_2}{y}\right) \Theta\left(-\frac{\eta_1\eta_3}{y}\right)\,\nonumber\\
	&=(-1)^{b+d-1}\frac{\delta(\zbar_{12})}{(z_{12})^{a+b+c+d-1}}
	\int ds \sum^\infty_{m,n=0} \frac{u^m_{12}}{m!}\frac{z^n_{12}}{n!} s^{m+n-b}(1-s)^{-c}\, \partial^m_{u_2}\partial^n_{z_2}\frac{\delta(z_{23})\delta(\zb_{23})}{u^d_{23}}\\
	&\qquad \qquad \qquad \qquad \qquad\qquad  \, \times \Theta\left(\frac{(1-y)\eta_1\eta_2}{y}\right) \Theta\left(-\frac{\eta_1\eta_3}{y}\right)\nonumber.
	\end{align}
	We recognize the appearance of the chiral OPE block \eqref{eq:chiralwitheta} by writing
	
	\begin{align}
	\langle O_1 O_2 O_3 \rangle=(-1)^{b+d-1}\delta(\zb_{12}) z^{h_4-h_1-h_2} &\int ds \sum^\infty_{m,n=0} \frac{u^m_{12}}{m!}\frac{z^n_{12}}{n!} s^{m+n-b}(1-s)^{-c}\langle P^m_{-1,-1} L^n_{-1} O_4 O_3\rangle \nonumber\\
	&   \times \Theta\left(\frac{(1-y)}{y}\eta_1\eta_2\right) \Theta\left(-\frac{\eta_1\eta_3}{y}\right)\,,
	\end{align}
	where the two-point function $\langle O_4 O_3\rangle$ is given by \eqref{2-point function 1}, and where the exchanged operator $O_4$ has quantum numbers 
	\begin{equation}
	h_4=-1-h_3+\hb_1+\hb_2+\hb_3\,, \qquad \hb_4=\hb_1+\hb_2-1\,,\qquad \eta_4=-\eta_3\,,
	\end{equation}
	or equivalently
	\begin{equation}
	\label{Delta4 J4}
	\Delta_4=\Delta_1+\Delta_2-J_1-J_2-J_3-2\,, \qquad J_4=-J_3\,,\qquad \eta_4=-\eta_3.
	\end{equation}
	As discussed below \eqref{eq:chiralwitheta}, the Heaviside functions determine the range of integration depending on the channel of the three-point function. In particular for the configuration $\eta_1=\eta_2=-\eta_3$, we find 
	\begin{equation}
	\label{3-point amplitude OPE result}
	\begin{split}
	&\langle O_1 O_2 O_3 \rangle\\
	&=(-1)^{b+d-1}\delta(\zb_{12}) z^{h_4-h_1-h_2}   \sum^\infty_{m,n=0} \frac{u^m_{12}}{m!}\frac{z^n_{12}}{n!} B(m+n+1-b,1-c)\langle P^m_{-1,-1} L^n_{-1} O_4 O_3\rangle.
	\end{split}
	\end{equation}
	The result for the other configurations only differ by an overall constant. The result \eqref{3-point amplitude OPE result} exactly agrees with \eqref{eq:deltaOPEansatz}-\eqref{eq:OPEcoefficientsdelta}.  
	
	We can go even further. In \eqref{eq:chiralOPEblock} we gave a formula for the chiral OPE block, and we can show that its contribution gives the full 3-point amplitude \eqref{3-point amplitude OPE}. Thus we set out to compute
	\begin{equation}
	\label{reconstruction formula}
	\langle O_1O_2O_3\rangle=\int d^3\x_4\, \frac{\delta(\zbar_{12})\delta(\zbar_{24})}{ (z_{12})^{\tilde a}\, (z_{24})^{\tilde b}\, (z_{14})^{\tilde c}\, (F_{124})^{\tilde d}}\, \langle O_4(\x_4) O_3(\x_3)\rangle\,,
	\end{equation}
	with $\tilde a,\tilde b,\tilde c,\tilde d$ given as in equations \eqref{d parameter}-\eqref{abc parameters} upon replacing $\Delta_3 \mapsto 3-\Delta_4$ and $J_3 \mapsto -J_4$, with $(\Delta_4,J_4)$ given by \eqref{Delta4 J4}, namely
	\begin{equation}
	\begin{split}
	\tilde a&=J_1+J_2+\Delta_4-1=\Delta_1+\Delta_2-J_3-3\,,\\
	\tilde b&=J_2-J_4-\Delta_1+2=J_2+J_3-\Delta_1+2=b\,,\\
	\tilde c&=J_1-J_4-\Delta_2+2=J_1+J_3-\Delta_2+2=c\,,\\
	\tilde d&=\Delta_1+\Delta_2-\Delta_4-J_1-J_2+J_4-1=1\,.
	\end{split}
	\end{equation}
	Inserting the relevant two-point function,
	\begin{equation}
	\langle O_4(\x_4) O_3(\x_3)\rangle=\frac{\delta(z_{34})\delta(\zbar_{34})}{(u_{34})^{\Delta_3+\Delta_4-2}}=\frac{\delta(z_{34})\delta(\zbar_{34})}{(u_{34})^{2(\hb_1+\hb_2+\hb_3-2)}}\,,
	\end{equation}
	we thus have
	\begin{equation}
	\langle O_1O_2O_3\rangle=\frac{\delta(\zbar_{12})\delta(\zbar_{23})}{ (z_{12})^{\tilde a}\, (z_{23})^{b}\, (z_{13})^{c}}\int   \frac{du_4}{(u_1 z_{23}+u_2 z_{31}+u_4 z_{12})\,(u_{34})^{2(\hb_1+\hb_2+\hb_3-2)}}\,.
	\end{equation}
	Now let us assume $2(\hb_1+\hb_2+\hb_3-2)=n+1$ with $n\in\mathbb{N}$, such that we can integrate by parts and use the residue theorem,
	\begin{equation}
	\begin{split}
	\langle O_1O_2O_3\rangle&=\frac{\delta(\zbar_{12})\delta(\zbar_{23})}{ (z_{12})^{\tilde a}\, (z_{23})^{b}\, (z_{13})^{c}}\int   \frac{du_4}{(u_1 z_{23}+u_2 z_{31}+u_4 z_{12})\,(u_{34})^{n+1}}\\
	&=\frac{\delta(\zbar_{12})\delta(\zbar_{23})}{ (z_{12})^{\tilde a-n}\, (z_{23})^{b}\, (z_{13})^{c}} \int   \frac{du_4}{(u_1 z_{23}+u_2 z_{31}+u_4 z_{12})^{n+1}\,u_{34}}\\
	&= \frac{2\pi i\, \delta(\zbar_{12})\delta(\zbar_{23})}{ (z_{12})^{\tilde a-n}\, (z_{23})^{b}\, (z_{13})^{c}\, (F_{123})^{n+1}}\,. 
	\end{split}
	\end{equation}
	We note that $d=n+1$ and $\tilde a-n=a$, such that we have reconstructed \eqref{3-point amplitude OPE} from the contribution of a single OPE block as encapsulated by \eqref{reconstruction formula}. It would be interesting to see how this computation generalizes to non-integer $d$. Note, however, that the condition $d=n+1$ is satisfied by the carrollian amplitudes \eqref{eq:3ptwithTheta} arising from carrollian primaries with $\Delta_i=1$ or their descendants.
	
	\subsection{4-point correlators and amplitudes} 
	\subsubsection*{Correlators}
	We start with the generic 4-point correlator given in \eqref{4-point general}. This time let us consider the coincidence limit
	\begin{equation}
	z_{12} \sim \zbar_{12}  \leq u_{12} \ll 1\,.
	\end{equation}
	The reason for demanding $z_{12} \sim \zbar_{12}$ comes from the presence of the delta distribution $\delta(z-\zbar)$, which requires to zoom into the region of vanishing $\zbar=z$. Indeed, as we take $z_{12} \to 0$ we have
	\begin{equation}
	z \stackrel{z_{12} \to 0}{\sim} z_{12}\, \frac{z_{34}}{z_{23}z_{24}} \,,
	\end{equation}
	and thus
	\begin{equation}
	\zbar_{12}=\frac{\zbar_{13} \zbar_{24}}{\zbar_{34}}\zbar \stackrel{z_{12}\to 0}{\sim} z_{12}\, \frac{z_{34} \zbar_{23} \zbar_{24}}{\zbar_{34} z_{23}z_{24}}\,.
	\end{equation}
	In that limit, we can write
	\begin{equation}
	F_{1234}\stackrel{z_{12}\to 0}{\sim} -\frac{u_{12}}{z_{12}} \frac{z_{24}\zbar_{34}}{\zbar_{23}}\,, \qquad \delta(z-\zbar)\stackrel{z_{12}\to 0}{\sim} \frac{\zbar_{23}\zbar_{24}}{\zbar_{34}} \delta(\zbar_{12})\,.
	\end{equation}
	To proceed we also assume a generic power-law behaviour for the undetermined function $G(z)$ appearing in the generic formula \eqref{4-point general}, 
	\begin{equation}
	G(z) \sim z^p\,, \qquad (z \sim 0)\,.
	\end{equation}
	Taken together, this yields
	\begin{equation}
	\label{4-point OPE limit}
	\langle O_1 O_2 O_3 O_4 \rangle \stackrel{z_{12}\to 0}{\sim} \frac{\delta(\zbar_{12})}{(-u_{12})^c (z_{12})^{b_{12}}} \frac{1}{(z_{23})^{b_{23}}(z_{24})^{b_{24}}(z_{34})^{b_{34}}(\zbar_{23})^{\bar b_{23}}(\zbar_{24})^{\bar b_{24}}(\zbar_{34})^{\bar b_{34}}}\,,
	\end{equation}
	with the exponents given by
	\begin{equation}
	\begin{split}
	b_{12}&\equiv a_{12}+\bar a_{12}-c-p\\
	&=\left(2\Delta_1+2\Delta_2-\Delta_3-\Delta_4-2 c-3p\right)/3\,,\\
	b_{23}&\equiv a_{13}+a_{23}-\bar a_{12}+p\\
	&=\left(h_1+h_2+4h_3-2h_4-2\bar h_1-2\bar h_2+\bar h_3+\bar h_4+c/2+3p \right)/3\\
	&=\left(-\Delta_1-\Delta_2+5\Delta_3-\Delta_4+3J_1+3J_2+3J_3-3J_4+c+6p \right)/6\,,\\
	b_{24}&\equiv a_{14}+a_{24}-\bar a_{12}+c+p\\
	&=\left(h_1+h_2-2h_3+4h_4-2\bar h_1-2\bar h_2+\bar h_3+\bar h_4+c/2+3p \right)/3\\
	&=\left(-\Delta_1-\Delta_2-\Delta_3+5\Delta_4+3J_1+3J_2-3J_3+3J_4+c +6p\right)/6\,,\\
	b_{34}&\equiv a_{34}+\bar a_{12}-p\\
	&=\left(-h_1-h_2+2h_3+2h_4+2\bar h_1+2\bar h_2-\bar h_3-\bar h_4-c/2-3p \right)/3\\
	&=\left(\Delta_1+\Delta_2+\Delta_3+\Delta_4-3J_1-3J_2+3J_3+3J_4-c-6p \right)/6\,,\\
	\bar b_{23}&\equiv \bar a_{12}+\bar a_{13}+\bar a_{23}-c-1\\
	&=\bar h_1+\bar h_2+\bar h_3-\bar h_4-c/2-1\\
	&=\left(\Delta_1+\Delta_2+\Delta_3-\Delta_4-J_1-J_2-J_3+J_4-c-2 \right)/2\,,\\
	\bar b_{24}&\equiv \bar a_{12}+\bar a_{14}+\bar a_{24}-1\\
	&=\bar h_1+\bar h_2-\bar h_3+\bar h_4-c/2-1\\
	&=\left(\Delta_1+\Delta_2-\Delta_3+\Delta_4-J_1-J_2+J_3-J_4-c-2\right)/2\,,\\
	\bar b_{34}&\equiv -\bar a_{12}+\bar a_{34}+c+1\\
	&=-\bar h_1-\bar h_2+\bar h_3+\bar h_4+c/2+1\\
	&=\left(-\Delta_1-\Delta_2+\Delta_3+\Delta_4+J_1+J_2-J_3-J_4+c +2\right)/2\,.
	\end{split}
	\end{equation}
	Provided the existence and validity of the carrollian OPE, the expression \eqref{4-point OPE limit} should take the form
	\begin{equation}
	\langle O_1(\x_1)O_2(\x_2)O_3(\x_3)O_4(\x_4) \rangle \sim f_{125}(\x_{12}) \langle O_5(\x_2) O_3(\x_3) O_4(\x_4) \rangle\,.
	\end{equation}
	We see that the second factor in \eqref{4-point OPE limit} takes the form of a time-independent three-point function, provided the conformal weights $(h_5,\bar h_5)$ of the exchanged operator $O_5$ satisfy
	\begin{equation}
	b_{23}=h_5+h_3-h_4\,, \qquad b_{24}=h_5+h_4-h_3\,, \qquad b_{34}=h_3+h_4-h_5\,,
	\end{equation}
	together with the conjugate relations. These constraints are solved at once by
	\begin{equation}
	\label{weights O5}
	\begin{split}
	h_5&=\left(h-2\bar h_1-2\bar h_2+\bar h_3+\bar h_4+c/2+3p \right)/3\,,\\
	\bar h_5&=\bar h_1+\bar h_2-c/2-1\,,
	\end{split}
	\end{equation}
	or equivalently
	\begin{equation}
	\label{dimensions O5}
	\begin{split}
	\Delta_5&=\frac{\Sigma \Delta-c}{3}+p-1\,,\\
	J_5&=\frac{-2\Delta_1-2\Delta_2+\Delta_3+\Delta_4+2c}{3}+J_1+J_2+p+1\,.
	\end{split}
	\end{equation}
	With these identifications the first factor in \eqref{4-point OPE limit} can be written  
	\begin{equation}
	f_{125}(\x_{12})=\delta(\zbar_{12}) (z_{12})^{J_5-J_1-J_2-1} (u_{12})^{2(\bar h_5-\bar h_1-\bar h_2+1)} \,,
	\end{equation}
	which can be recognized as one of the structure functions in \eqref{f12k}.
	
	\subsubsection*{Amplitudes}
	Consider the most general from of the carrollian 4-point amplitude with the Heaviside functions coming from energy positivity, which we reproduce here for convenience
	\begin{align}
	\label{4-point start}
	\langle O_1O_2O_3O_4\rangle&= \delta(z-\zb) G(z)\prod_{i<j}\frac{1}{(z_{ij})^{a_{ij}}(\zbar_{ij})^{\bar{a}_{ij}}(F_{1234})^{c}} \\
	&\times \Theta\left(-z\left|\frac{z_{24}}{z_{12}}\right|^2 \eta_1 \eta_4  \right)\Theta\left(\frac{1-z}{z} \left|\frac{z_{34}}{z_{23}}\right|^2\eta_2 \eta_4  \right)\Theta\left(-\frac{1}{1-z}\left|\frac{z_{14}}{z_{13}}\right|^2\eta_3 \eta_4 \right).\nonumber
	\end{align}
	The constants $a_{ij},\bar{a}_{ij}$ are determined in \eqref{aij} while $c$ and $G(z)$ are not fixed by symmetries. On the support of the latter, we can rewrite the $\Theta$ functions as
	\begin{equation}
	\label{step functions}
	\begin{split}
	\Theta\left(-z\left|\frac{z_{24}}{z_{12}}\right|^2 \eta_1 \eta_4  \right)&=\Theta\left(-\eta_1\eta_4 \frac{z_{34}\zb_{24}}{z_{13}\zb_{12}}\right)=\Theta\left(\eta_5\eta_4 \frac{z_{34}}{z_{13}}\right)\Theta\left(-\eta_1\eta_5 \frac{\zb_{24}}{\zb_{12}}\right)\,,\\
	\Theta\left(\frac{1-z}{z} \left|\frac{z_{34}}{z_{23}}\right|^2\eta_2 \eta_4  \right)&=\Theta\left(\eta_6\eta_4\frac{z_{34}}{z_{23}}\right)\Theta\left(\eta_2\eta_6\frac{\zb_{13}}{\zb_{12}}\right)\,,\\
	\Theta\left(-\frac{1}{1-z}\left|\frac{z_{14}}{z_{13}}\right|^2\eta_3 \eta_4 \right)&=\Theta\left(-\eta_3\eta_4\eta_7\frac{z_{24}}{z_{23}}\right)\Theta\left(\eta_7\frac{\zb_{14}}{\zb_{13}}\right)\,,
	\end{split}
	\end{equation}
	where we introduced the additional in/out labels $\eta_{5,6,7}=\pm 1$ in order to split the step functions. 
	
	We will consider the OPE limit $z_{12}\,,\zbar_{12} \sim 0$. We note that the result will depend on the order of limits so that we will always assume the consecutive limits $z_{12}\rightarrow 0$ followed by $\zb_{12}\rightarrow 0$,
	It will be convenient to eliminate $\zb_{34}$ using the delta distribution,\footnote{We are thus focusing on the contribution from one kinematic region within the support of the distribution.} 
	\begin{equation}
	\label{eq:7}
	\delta(z-\zb)\sim \delta(\zb_{34}) \textrm{sgn}(\zb_{12}\zb_{13} \zb_{24}) \frac{\zb_{13}\zb_{24}}{\zb_{12}}.
	\end{equation}
	We see therefore that we have to set $\eta_7=1$ and $\eta_5=\eta_6$ in order to have a non-zero result for the step functions \eqref{step functions} in this limit. Assuming the behavior $G(z)\sim z^p$ for $z\sim 0$ as usual, in the OPE limit the 4-point function \eqref{4-point start} can be written as
	\begin{align}
	\nonumber
	\langle O_1O_2O_3O_4\rangle &\overset{z_{12}\to 0}{\sim} \delta(\zb_{34}) \textrm{sgn}(\zb_{12}\zb_{13} \zb_{23})(z_{12})^{-a_{12}+p} (z_{23})^{-a_{13}-a_{23}-p} (z_{24})^{-a_{14}-a_{24}-p} (z_{34})^{-a_{34}+p} \\
	&\times (\zb_{12})^{-\bar{a}_{12}+\bar{a}_{34}-1} (\zb_{12}+\zb_{23})^{-\bar{a}_{13}-\bar{a}_{14}-\bar{a}_{34}+1} (\zb_{23})^{-\bar{a}_{23}-\bar{a}_{24}-\bar{a}_{34}+1} (F_{1234})^{-c} \label{eq:first4ptexp}\\
	&\times \Theta\left(\eta_5\eta_4\frac{z_{34}}{z_{23}}\right)\Theta\left(-\eta_3\eta_4\frac{z_{24}}{z_{23}}\right)\Theta\left(-\eta_1\eta_5\frac{\zb_{23}}{\zb_{12}}\right)\Theta\left(\eta_2\eta_5 \frac{\zb_{12}+\zb_{23}}{\zb_{12}}\right)\nonumber.
	\end{align}
	Note that in this limit we also have
	\begin{equation}
	z_{23}F_{1234}\sim z_{23}u_4+u_3 z_{42}+z_{34}\frac{\zb_{12}+\zb_{23}}{\zb_{12}}\left(u_2-u_1 \frac{\zb_{23}}{(\zb_{12}+\zb_{23})}\right).
	\end{equation}
	As in the case of the 3-point function, inspection of the Heaviside functions shows that we have to zoom in on the kinematic region $\zb_{23}\sim 0$ in the limit $\zb_{12}\rightarrow 0$ for a nonzero result. As before, we will treat the correlator $\langle O_1O_2O_3O_4\rangle$ as a distribution in $\zb_{23}$ where we expect to have an emergent delta function. To make this explicit, we use the same trick as in \eqref{eq:3pttrick} and introduce unity in terms of an integral over a delta function $\delta(x-\zb_{23})$ on the right-hand side of \eqref{eq:first4ptexp}. Changing the integration variables to $x=-t \zb_{12}$, we obtain
	\begin{align}
	&\langle O_1O_2O_3O_4\rangle \overset{z_{12}\rightarrow 0}{\sim }  \sgn(\eta_1\eta_2\zb_{12})(z_{12})^{-a_{12}+p} (\zb_{12})^{-\Sigma\, \bar a_{ij}+1} \nonumber \\
	&\times \delta(\zb_{34}) \Theta\left(\eta_5\eta_4\frac{z_{34}}{z_{23}}\right)\Theta\left(-\eta_3\eta_4\frac{z_{24}}{z_{23}}\right)(z_{23})^{-a_{13}-a_{23}-p+c} (z_{24})^{-a_{14}-a_{24}-p} (z_{34})^{-a_{34}+p} \label{eq:4to3final}\\
	&\times \int dt \frac{\delta(\zb_{23}+t \zb_{12})(1-t)^{-\bar{a}_{13}-\bar{a}_{14}-\bar{a}_{34}+1} (-t)^{-\bar{a}_{23}-\bar{a}_{24}-\bar{a}_{34}+1}}{\left(z_{24}u_4+u_3 z_{42}+z_{34}(u_2+t u_{12})\right)^c}\Theta(\eta_1\eta_5 t)\Theta(\eta_2\eta_5(1-t)),\nonumber
	\end{align}
	where we wrote $\sgn(\zb_{12}\zb_{13}\zb_{23})=\sgn (\zb_{12}t (1+t))=\sgn(-\zb_{12}\eta_1\eta_2)$ on account of the Heaviside functions.
	Expanding in powers of $u_{12}, \zb_{12}$ we can recognize this as a sum over the carrollian 3-point function $\langle O_5O_4O_3\rangle$ and its derivatives, where $O_5$ has weights
	\begin{equation}
	\begin{split}
	h_5&=\frac{1}{3}(h_1+\hb_1+h_2+\hb_2+h_3-2\hb_3+h_4-2\hb_4)+p+\frac{c}{6}\,,\\
	\hb_5&=2-\hb_3-\hb_4+\frac{c}{2}\,.
	\end{split}
	\end{equation}
	More explicitly, we can write
	\begin{equation}
	\begin{split}
	&\langle O_1O_2O_3O_4\rangle \overset{z_{12}\rightarrow 0}{\rightarrow }  \sgn(-\eta_1\eta_2\zb_{12}) (-1)^{\hb_5-\hb_2+\hb_1}(z_{12})^{h_5-h_1-h_2}  (\zb_{12})^{\hb_5-\hb_1-\hb_2} \\
	&   \qquad  \int dt\,  \Theta(\eta_1\eta_5 t)\Theta(\eta_2\eta_5(1-t)) \sum^\infty_{m,n=0} \frac{u^m_{12}}{m!}\frac{\zb^n_{12}}{n!} 
	(1-t)^{\hb_5+\hb_2-\hb_1-1} t^{\hb_5-\hb_2+\hb_1+m+n-1}
	\\
	&\hspace{7cm} \times\langle P^m_{-1,-1}\bar{L}^n_{-1} O_5O_4O_3\rangle\,.
	\end{split}
	\end{equation}
	In case that both $O_1,O_2$ and the exchanged operator $O_5$ are all in/outgoing ($\eta_1=\eta_2=\eta_5$) the integration range is $t\in(0,1)$ and we recognise immediately the holomorphic OPE expansion~\eqref{eq:holomorphicOPE}. For the other in/out configurations, the resulting expansion is the same up to an overall coefficient.
	
	%In the case $\eta_1=-\eta_2$, the parameter $\eta_5$ is not fixed. Setting $\eta_5=\eta_1$, one has $t<-1$ and the coefficient in the second line of \eqref{eq:OPE3to4} is replaced by $(-1)^{-\bar{a}_{13}-\bar{a}_{14}-\bar{a}_{23}-\bar{a}_{24}} B(1-\bar{a}_{13}-\bar{a}_{14},-1+\bar{a}_{13}+\bar{a}_{14}+\bar{a}_{23}+\bar{a}_{24})$. On the other hand, for $\eta_1=-\eta_2=-\eta_5$, one has $t>0$ with resulting coefficient $B(1-\bar{a}_{23}-\bar{a}_{24},-1+\bar{a}_{13}+\bar{a}_{14}+\bar{a}_{23}+\bar{a}_{24})$.
	
	In conclusion, the examples worked out in section~\ref{section 6} give substantial evidence that the carrollian OPEs constructed in section~\ref{section 5} control the short-distance expansion of carrollian correlators and amplitudes. We emphasise that the structures uncovered here go beyond that resulting from the well-known colinear factorisation of momentum amplitudes. Indeed, while the latter is encoded in the so-called `holomorphic OPE', we have found that other carrollian OPE branches control short-distance expansions of carrollian amplitudes, even for the 4-point contact scalar amplitude where colinear factorisation does not apply. This opens up new ways to study and constrain carrollian amplitudes, that are similar in spirit to the standard conformal bootstrap. The development of this carrollian toolbox, and its application to the study of massless scattering amplitudes, will be the subject of future works.

	\section*{Acknowledgments}
	We thank Tim Adamo and Sabrina Pasterski for stimulating discussions. The work of KN and JS is supported by two Postdoctoral
	Research Fellowships granted by the F.R.S.-FNRS (Belgium).

	\bibliography{bibl}
	\bibliographystyle{JHEP}  
\end{document}